\documentclass{egpubl}
\usepackage{eg2025}
 
\ConferencePaper        
\CGFccby

\usepackage[T1]{fontenc}
\usepackage{dfadobe}  
\usepackage{amsmath}
\usepackage{bm}
\usepackage{hyperref}
\usepackage{booktabs}
\usepackage{subcaption}
\usepackage{multirow}
\usepackage{siunitx}
\usepackage{microtype}
\usepackage{xspace}
\usepackage{amssymb}
\usepackage{xcolor}

\usepackage{cite}  
\BibtexOrBiblatex
\electronicVersion
\ifpdf \usepackage[pdftex]{graphicx} \pdfcompresslevel=9
\else \usepackage[dvips]{graphicx} \fi

\usepackage{egweblnk} 

\title[Corotational Hinge-based Thin Plates/Shells]%
      {Corotational Hinge-based Thin Plates/Shells}

\author[Qixin Liang]
{
        \parbox{\textwidth}{\centering Qixin Liang\orcid{0009-0002-5494-3672}}
        \\
        {\parbox{\textwidth}{\centering The University of Hong Kong \& TransGP, Hong Kong}
}
}

\begin{document}


\maketitle
\begin{abstract}
   We present six thin plate/shell models, derived from three distinct types of curvature operators formulated within the corotational frame, for simulating both rest-flat and rest-curved triangular meshes. Each curvature operator derives a curvature expression corresponding to both a plate model and a shell model. The corotational edge-based hinge model uses an edge-based stencil to compute directional curvature, while the corotational FVM hinge model utilizes a triangle-centered stencil, applying the finite volume method (FVM) to superposition directional curvatures across edges, yielding a generalized curvature. The corotational smoothed hinge model also employs a triangle-centered stencil but transforms directional curvatures into a generalized curvature based on a quadratic surface fit. All models assume small strain and small curvature, leading to constant bending energy Hessians, which benefit implicit integrators. Through quantitative benchmarks and qualitative elastodynamic simulations with large time steps, we demonstrate the accuracy, efficiency, and stability of these models. Our contributions enhance the thin plate/shell library for use in both computer graphics and engineering applications.


\end{abstract} 
\section{Introduction}

Thin shells based on Kirchhoff–Love (KL) theory account for both membrane and bending components, and when a thin shell with a rest-curved configuration is relaxed to a rest-flat shape when unstressed, it is treated as a thin plate. Compared to the membrane component, the bending component is particularly important for capturing the formation of folds and wrinkles. In this study, we concentrate on developing and analyzing bending models for thin plates and shells.

Modeling and simulating thin plates/shells is a longstanding topic in both the computer graphics community and computational mechanics community.
The computational mechanics community has developed numerous bending formulations for plates and shells, with a strong emphasis on accuracy. Examples include methods based on subdivision surfaces~\cite{cirak2000subdivision} and Isogeometric Analysis (IGA) with high-order continuity solution spaces~\cite{benson2011large}, which require sophisticated approaches to apply boundary conditions and handle contact resolution on control nodes. In contrast, the computer graphics community prioritizes efficiency, favoring models such as the hinge angle-based bending model~\cite{BW98, grinspun2003discrete, Bridson2003Cloth}, Quadratic Shells~\cite{Bergou2006Aquadratic}, and Cubic Shells~\cite{Garg2007cubicshell} for their simplicity. However, these discretized edge-based energies have no bending along the common hinge edges, leading to a failure to converge to the complete shape operator of a smooth surface at element interfaces.
Recently, the pursuit of an efficient and accurate algorithm has been a common goal for both communities~\cite{Chen2018Physical, weischedel2012discrete, sauer2024simple}. In this study, we take a step further at this intersection.

To summarize, our main contributions are as follows:
\begin{itemize}
    \item We present a \textit{corotational edge-based hinge curvature operator} for thin shell simulation, including a specific variant for rest-flat (thin plate) simulation.
    
    \item We propose a \textit{corotational FVM (finite volume method) hinge curvature operator} for thin shell simulation, along with a specific variant for rest-flat (thin plate) simulation.
    
    \item We introduce a \textit{corotational smoothed hinge curvature operator} for thin shell simulation, as well as a specific variant for rest-flat (thin plate) simulation.

    \item All models feature constant bending energy Hessians, with detailed boundary conditions for accurate simulation.
\end{itemize}

Our models based on the linear triangle mesh structure are easy to integrate into existing finite element codes and thin shell (cloth) simulators.

\section{Related Works}
\subsection{Computational models of thin plates and shells}

Computational modeling and simulation of bending behaviors in thin flexible objects is an active research topic in the graphics community. In 1987, the seminal work~\cite{Terzopoulos1987Elastically} introduced tensorial treatments of the second fundamental form, discretized using the finite difference method on a regular quadrilateral grid, to model cloth and deformed surfaces. Subsequently, particle and mass-spring methods~\cite{Breen1994Predicting, Volino1995, Choi2002Stable} were explored as alternatives to these complex tensorial treatments, aiming to improve efficiency. However, these methods often compromised physical accuracy, resulting in material parameters that were mesh-dependent and not easily transferable across different mesh topologies.
Baraff and Witkin~\cite{BW98} utilized the hinge angle to model the bending constraint, similar to the approach in~\cite{Volino1995}, but they focused on the rest-flat configuration for cloth modeling. The hinge angle is measured on an edge-based hinge stencil, which composes a hinge edge and its two adjacent triangles. Building on this stencil, Grinspun et al.~\cite{grinspun2003discrete} described the bending energy of the rest-curved thin shell on a discrete differential geometry view. In the same published volume, Bridson et al.~\cite{Bridson2003Cloth} offered a productive formulation of the hinge bending model. When the curvature is small, these two models can be related by scaling a coefficient~\cite{Romero2021test, Feng2022Learning}. Both models have been adopted by well-known cloth simulators, C-IPC~\cite{Li2021CIPC} and Arcsim~\cite{Narain2012arcsim}, respectively, due to their simplicity. However, the edge-based hinge bending model is limited in its ability to capture complete local curvature behavior and suffers from mesh dependency~\cite{Grinspun2006ComputingDS}.

To enhance the consistency of the edge-based hinge bending model, a hinge-averaged shape operator~\cite{gingold2004discrete} was described to model the bending strain, which has been applied to simulate plasticity and fracture~\cite{Pfaff2014tearing}. Despite this improvement, the convergence to the ground truth solution remains slow~\cite{Chen2018Physical, Grinspun2006ComputingDS}. This issue was effectively addressed by introducing an additional degree of freedom (DoF) for midedge normal rotation to correct the hinge-averaged shape operator, resulting in what is known as the midedge operator~\cite{Zorin2005Curvature, Grinspun2006ComputingDS}. This operator has been integrated into libshell~\cite{Chen2018Physical}, which is based on the shear-rigid Koiter shell model, and is also included in the shearable Cosserat shell model~\cite{weischedel2012discrete}. While the midedge operator improves consistency and convergence in thin-shell simulations, the extra midedge DoFs introduce a higher computational burden. In our models, the extra midedge DoFs are not involved.

To improve the efficiency of the edge-based hinge bending model, the hinge angle is linearized under the quasi-isometry (small in-plane strain) condition, resulting in a constant bending energy Hessian and a linear bending force for rest-flat thin shells, known as the Quadratic Shell model~\cite{Bergou2006Aquadratic, Bergou2006SigCourse}. This model can be generalized from the Cubic Shell model~\cite{Garg2007cubicshell}, which is suitable for rest-curved thin shell configurations. Although both models offer improved efficiency, the inherent drawbacks of the edge-based hinge bending model~\cite{grinspun2003discrete} persist. The quadratic biharmonic energy~\cite{WARDETZKY2007} has a triangle-centered stencil and tends to perform better in terms of convergence to ground truth than models on the edge-based stencil, though it is limited to isometric, pure bending of plates. Based on the triangle-centered stencil, we introduce an FVM hinge curvature operator and a smoothed hinge curvature operator derived in the corotational frame~\cite{CrisfieldFEM1997} to address the limitations of the abovementioned edge-based models.

More recently, Le et al.~\cite{Le2023Second-Order} proposed a second-order Discrete Shells~\cite{grinspun2003discrete} model, demonstrating superior efficiency from the second-order triangle. Similarly benefited from the second-order tessellation, Löschner et al.~\cite{loschner2024curved} showcased the effectiveness of a second-order three-director finite element shell with microrotation fields. However, computational modeling with second-order elements requires abandoning the piecewise linear triangle structure. Wen and Barbi\v{c}~\cite{Wen2023KLShell} focused on deriving the KL thin-shell mechanical energy for arbitrary 3D volumetric hyperelastic materials, building their computational model from the foundation in the discrete geometry shell~\cite{Chen2018Physical, weischedel2012discrete}.

Another relevant topic in the computational mechanics community is the concept of a "rotation-free shell", where the shell is characterized without nodal rotational DoF. For those interested, a comparison study by G{\"a}rdsback et al.~\cite{gardsback2007comparison} provides insight, although it focuses on linear shell analysis. More recently, Zhou et al.~\cite{zhou2012geometric} extended a rotation-free beam model to a rotation-free shell model. While this approach offers certain efficiency advantages, its complex and laborious boundary condition treatment limits its potential applications. The pursuit of high-fidelity and high-performance (accurate, low computational cost, robustness, low sensitivity to poorly shaped meshes) thin-shell simulations increasingly blurs the lines between different research communities. Our models mainly draw inspiration from both discrete geometry shells~\cite{Bergou2006Aquadratic, Grinspun2006SigCourse, Garg2007cubicshell, Grinspun2006ComputingDS, weischedel2012discrete} and rotation-free shells~\cite{onate2000rotation, zhou2012geometric}.

\subsection{Corotational approach} 

To maintain the convergence properties of a linear approach while accommodating arbitrarily large rigid body transformations, the corotational approach is commonly employed to measure pure deformation. M\"{u}ller et al.~\cite{Muller2002Stable} first introduced the corotational formulation to handle the geometric non-linearity for deformed body simulations in graphics community. Since then, this formulation has been widely adopted in computer graphics applications for stable and efficient solid simulations~\cite{Zhu2010multigrid, Kugelstadt2018fast}.

In thin shell formulations with the corotational approach, Etzmuß et al.~\cite{Etzmu2003fast} extract the rotational component from the deformation gradient for each element and apply it to compute the bending stiffness matrices. In a similar vein to extract the rotation field, Thomaszewski et al.~\cite{Thomaszewski2006subdiv} adopt subdivision basis functions~\cite{cirak2000subdivision} to improve accuracy in cloth simulation, but this basis function comes with significant computational costs. More recently, a smoothed hinge model~\cite{liang2024smoothed} based on the corotational formulation has been developed to measure pure deformation in the corotational frame, which undergoes only rigid-body motion, for cloth simulation. However, this model is limited to the rest-flat thin shell configuration. In our work, we provide all formulations by providing curvature operators for both rest-flat and -curved configurations, along with detailed boundary treatments. Leveraging the corotational approach, all our models feature constant bending energy Hessians, which enhances the efficiency and stability for implicit simulations.

\section{Thin Shell Mechanics}
In a KL thin shell, the elastic shell energy $\Psi_\mathit{shell}$ is composed of both membrane energy $\Psi_m$ and bending energy $\Psi_b$:
\begin{equation} \label{eq1}
    \Psi_\mathit{shell} = \Psi_m + \Psi_b = 
    \frac{1}{2} \int_{\bar{\Omega}} \boldsymbol{\varepsilon}_\mathit{m}^T \mathbf{D}_\mathit{m} \boldsymbol{\varepsilon}_\mathit{m} d \bar{\Omega} + \frac{1}{2} \int_{\bar{\Omega}} \boldsymbol{\varepsilon}_\mathit{b}^T  \mathbf{D}_\mathit{b} \boldsymbol{\varepsilon}_\mathit{b} d \bar{\Omega},
\end{equation}
where ${\bar{\Omega}}$ represents the initial configuration of the shell's mid-surface. The vectors $\boldsymbol{\varepsilon}_{\mathit{m}}$ and $\boldsymbol{\varepsilon}_{\mathit{b}}$ respectively denote the membrane strain and curvature change expressed in Voigt notation. The membrane stiffness matrix $\mathbf{D}_{\mathit{m}} = h\mathbf{E}$ and the bending stiffness matrix $\mathbf{D}_{\mathit{b}} = {h^3 \mathbf{E}}/{12}$ depend on the shell thickness $h$ and the elastic matrix $\mathbf{E}$, derived from the Saint Venant–Kirchhoff model. 

In this study, we primarily focus on the computational modeling of the bending component, and therefore, we adopt the constant strain triangle as the computational model for the membrane part. The curvature change $\boldsymbol{\varepsilon}_\mathit{b}$ reflects the change in curvature from the initial configuration $\boldsymbol{\kappa}_{0}$ to the current configuration $\boldsymbol{\kappa}$, i.e.,
\begin{equation} \label{eq2}
    \boldsymbol{\varepsilon}_\mathit{b} = \boldsymbol{\kappa} - \boldsymbol{\kappa}_{0}.
\end{equation}
In cases where the shell is initially flat, the initial curvature $\boldsymbol{\kappa}_{0}$ vanishes, reducing the model to that of a thin plate.

\section{Geometric Discretization}
\subsection{Kinematics}
The shell stencil in the corotational edge-based hinge bending model is defined by an edge-based stencil, consisting of one edge and its two adjacent triangles (see Figure~\ref{fig:corotational_edge_based_hinge}). On the other hand, the shell stencil for the corotational FVM/smoothed hinge bending model uses a triangle-centered stencil, which includes one central triangle and its three neighboring triangles (see Figure~\ref{fig:corotational_FVM_hinge} and Figure~\ref{fig:corotational_smoothed_hinge}). For a given point $\mathbf{x} \in R^{3}$ within one shell stencil, the position in the current configuration is computed as:
\begin{equation} \label{eq3}
    \mathbf{x} = \mathbf{X} + \mathbf{u},
\end{equation}
where $\mathbf{X} \in R^{3}$ is the position in the initial configuration, and $\mathbf{u} \in R^{3}$ is the displacement.

\subsection{Terminologies and Remarks}
\textit{Terminologies.} In this context, $(\cdot)^{{e}}$ denotes a vector that collects the quantities associated with a shell stencil. For instance, ${\mathbf{x}}^{e} = \begin{bmatrix}{\mathbf{x}}_{1}^{T}&{\mathbf{x}}_{2}^{T}&{\mathbf{x}}_{3}^{T}&{\mathbf{x}}_{4}^{T}\end{bmatrix}^{T}$ aggregates the current nodal positions for an edge-based stencil, while ${\mathbf{x}}^{e} = \begin{bmatrix}{\mathbf{x}}_{1}^{T}&{\mathbf{x}}_{2}^{T}&{\mathbf{x}}_{3}^{T}&{\mathbf{x}}_{4}^{T}&{\mathbf{x}}_{5}^{T}&{\mathbf{x}}_{6}^{T}\end{bmatrix}^{T}$ aggregates the current nodal positions for a triangle-centered stencil. 
The notation $(\cdot)_{0}$ denotes the quantity in the initial configuration. The notation $(\cdot)_{ij}$ denotes a directed line segment from point $(\cdot)_{i}$ to point $(\cdot)_{j}$ in any coordinate frame.
The tilde $\tilde{(\cdot)}$ represents the quantity defined in the corotational frame ($\tilde{X}-\tilde{Y}-\tilde{Z}$ is the initial corotational frame and $\tilde{x}-\tilde{y}-\tilde{z}$ is the current corotational frame). Quantities in the corotational frame can be transformed from those in the world frame, as illustrated in Appendix~\ref{appendixA}. 
$(\cdot)_{{c}}$ denotes projection.
More graphical illustration can be found in the accompanying Figure~\ref{fig:corotational_edge_based_hinge}, Figure~\ref{fig:corotational_FVM_hinge} and Figure~\ref{fig:corotational_smoothed_hinge}. 

The terms "EP", "ES", "FP", "FS", "SP", and "SS" are used to distinguish between different bending formulations, when necessary. Specifically, "E" refers to the edge-based hinge, "F" to the FVM hinge, and "S" to the smoothed hinge. "P" indicates a plate (rest-flat configuration), while "S" denotes a shell (rest-curved configuration). 

\textit{Remarks.} Transitioning from simpler cases to more complex scenarios, we provide more derivation details for the bending models based on edge-centered stencils using corotational approach, enabling a seamless generalization of this derivation process to bending models on triangle-centered stencils.
To facilitate understanding of the derivation process, we outline several key points.

The kinematics of a shell stencil deformed from its initial configuration to the current configuration can be expressed as $\mathbf{x}^{e}= \mathbf{X}^{e}+\mathbf{u}^{e}$. In the initial configuration, a shell stencil with ${\mathbf{X}}^{e}$ projected onto the tangential plane of the initial corotational frame is referred to as a corotational shell stencil with ${\mathbf{X}}^{e}_{c}$. By projecting the deviation vector $\mathbf{X}^{e}-\mathbf{X}^{e}_{c}$ along the normal direction of the initial corotational frames, and introducing curvature operators constructed from ${\mathbf{X}}^{e}_{c}$, the discretized curvature in the initial configuration can be defined. 
Under the small (in-plane) strain and curvature assumption, the relative positions of the projected positions ${\mathbf{x}}^{e}_{c}$ in the current corotational frame and ${\mathbf{X}}^{e}_{c}$ in the initial corotational frame are approximately identical. (Another view is that the corotational shell stencil transitions from the initial configuration to the current configuration as \({\mathbf{x}}^{e}_{c} = {\mathbf{X}}^{e}_{c} + {\mathbf{u}}^{e}_{c}\), approximately undergoing rigid-body motion). So, the discretized curvature in the current configuration can be defined using the curvature operators of the discretized curvature in the initial configuration. The bending deformation is quantified by the change in curvature between the initial and current configurations. Based on this, the constant bending energy Hessians can be rationally derived. Further details are provided in the subsequent subsections. Our source code (\url{https://github.com/liangqx-hku/libThinPlateShells}) is also made available to support practitioners.

\subsection{Corotational edge-based hinge bending model} \label{Corotational_edge-based_hinge_bending_model}

\begin{figure}
  \centering
  \includegraphics[width=.99\linewidth]{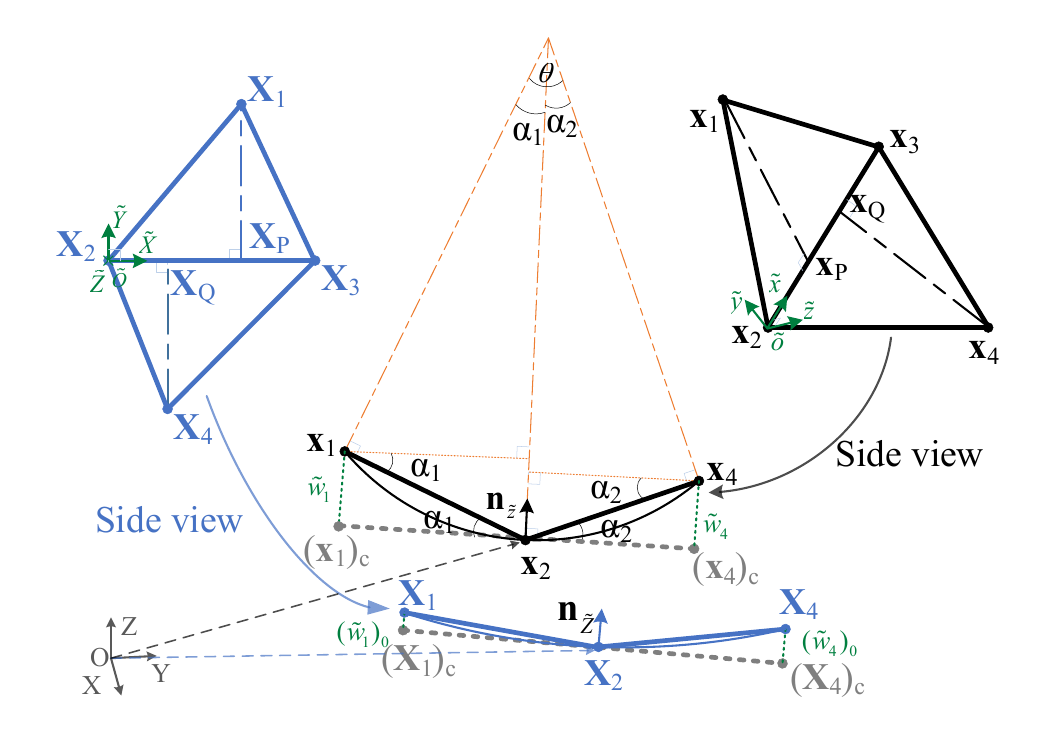}
  \caption{\label{fig:corotational_edge_based_hinge}
           A shell stencil (edge-based hinge) deforms from its initial configuration (blue) to the current configuration (black). The corotational frame (green) is employed to capture nodal deviations relative to the corotational shell stencil (gray dashed line). 
           $X-Y-Z$ is the world frame. For the shell stencil in the current configuration, two views are presented. $\mathbf{x}_{P}$ and $\mathbf{x}_{Q}$ are perpendicular feet. The $\tilde{x}-\tilde{y}-\tilde{z}$ frame represents the current corotational frame, where the $\tilde{x}$-axis aligns with the edge $\mathbf{x}_{23}$, and the $\tilde{z}$-axis direction (the direction of $\mathbf{n}_{\tilde{z}}$) is approximately along the bisector of the bend angle $\theta$. The $\tilde{y}$-axis is determined by the right-hand rule. The corotational shell stencil is the projection of the shell stencil onto the $\tilde{x}-\tilde{y}$ plane. $\alpha_{1}$ and $\alpha_{4}$ describe the bend angles of two adjacent triangles relative to the corotational shell stencil. The point $(\mathbf{x}_1)_c$ represents the projection of $\mathbf{x}_1$ onto the $\tilde{x}-\tilde{y}$ plane, while the transverse displacement $\tilde{w}_{1}$ corresponds to the projection of the relative difference between $\mathbf{x}_1$ and $(\mathbf{x}_1)_c$ along the $\tilde{z}$-axis. Similar quantities for other points can be calculated in the same manner. Variables related to the shell stencil in the initial configuration can be extended from those in the current configuration.}
           
\end{figure}

For an edge-based stencil based on the triangular mesh (see Figure~\ref{fig:corotational_edge_based_hinge}), there is no bending along the hinge edge. In the current configuration, the bend angle $\theta$ is small under the small strain/curvature assumption. The directional curvature $\kappa_{{m}}$ across the hinge edge in the current configuration can be expressed as
\begin{equation} \label{eq4}
    \kappa_{{m}} = \frac{2sin\theta}{\sqrt{h_1^{2} + 2h_1h_4\cos\theta + h_4^{2}}} \simeq \frac{2\theta}{h_1 + h_4}
\end{equation}
with $\theta \simeq \sin\theta$ and $\cos\theta \simeq 1$. The heights $h_1$ and $h_4$ correspond to the triangle $T_{123}$ and $T_{432}$, respectively, as approximated in the initial configuration. Under the small curvature assumption, the bend angle in the current configuration can be approximated by
\begin{equation} \label{eq5}
    \theta=\alpha_{1} + \alpha_{2} 
    \simeq  \sin\alpha_{1}+\sin\alpha_{2}
    \simeq \frac{\mathbf{n}_{\tilde{z}}^{T}({\tilde{\mathbf{x}}}_{1}-{\tilde{\mathbf{x}}}_{P})}{h_{1}}+\frac{\mathbf{n}_{\tilde{z}}^{T}({\tilde{\mathbf{x}}}_{4}-{\tilde{\mathbf{x}}}_{Q})}{h_{4}}.
\end{equation}
Here, the perpendicular feet $\tilde{\mathbf{x}}_{P}$ and $\tilde{\mathbf{x}}_{Q}$ are given by
\begin{equation} \label{eq6}
    \tilde{\mathbf{x}}_{P} = \frac{\lVert \tilde{\mathbf{x}}_{P3}\rVert}{\lVert \tilde{\mathbf{x}}_{23}\rVert}\tilde{\mathbf{x}}_{2}+\frac{\lVert \tilde{\mathbf{x}}_{P2}\rVert}{\lVert \tilde{\mathbf{x}}_{23}\rVert}\tilde{\mathbf{x}}_{3} , \ \tilde{\mathbf{x}}_{Q} = \frac{\lVert \tilde{\mathbf{x}}_{Q3}\rVert}{\lVert \tilde{\mathbf{x}}_{23}\rVert}\tilde{\mathbf{x}}_{2}+\frac{\lVert \tilde{\mathbf{x}}_{Q2}\rVert}{\lVert \tilde{\mathbf{x}}_{23}\rVert}\tilde{\mathbf{x}}_{3}.
\end{equation}
Under the small strain assumption, these perpendicular feet in Eq. (\ref{eq6}) can be approximated as
\begin{equation} \label{eq7}
    \tilde{\mathbf{x}}_{P} = \frac{\lVert \tilde{\mathbf{X}}_{P3}\rVert}{\lVert \tilde{\mathbf{X}}_{23}\rVert}\tilde{\mathbf{x}}_{2}+\frac{\lVert \tilde{\mathbf{X}}_{P2}\rVert}{\lVert \tilde{\mathbf{X}}_{23}\rVert}\tilde{\mathbf{x}}_{3} , \ \tilde{\mathbf{x}}_{Q} = \frac{\lVert \tilde{\mathbf{X}}_{Q3}\rVert}{\lVert \tilde{\mathbf{X}}_{23}\rVert}\tilde{\mathbf{x}}_{2}+\frac{\lVert \tilde{\mathbf{X}}_{Q2}\rVert}{\lVert \tilde{\mathbf{X}}_{23}\rVert}\tilde{\mathbf{x}}_{3},
\end{equation}
and $\mathbf{n}_{\tilde{z}}=\mathbf{n}_{c}/\lVert \mathbf{n}_{c} \rVert$ is the direction of the $\tilde{z}$-axis, with $\mathbf{n}_{c}$ expressed as
\begin{equation} \label{eq8}
    \mathbf{n}_{c}=\frac{\mathbf{x}_{P1}}{ \lVert\mathbf{x}_{P1}\rVert}+\frac{\mathbf{x}_{Q4}}{ \lVert\mathbf{x}_{Q4}\rVert}.
\end{equation}
It should be noted that we use the direction $\mathbf{n}_{\tilde{z}}$, which bisects the supplementary angle $(\pi - \theta)$ of the bend angle, approximates the normal direction of the smooth shell midsurface in the current configuration. 
To avoid the numerical issue, when $T_{123}$ and $T_{432}$ are coplanar, $\mathbf{n}_{\tilde{z}}$ aligns with the normal of triangle $T_{123}$ ($T_{432}$).

By substituting Eq. (\ref{eq7}) and Eq. (\ref{eq5}) into Eq. (\ref{eq4}), the directional curvature in the current configuration is discretized as
\begin{equation} \label{eq9}
    \kappa_{m} = {\mathbf{L}}_{m}\mathbf{N}_{}^{T}\tilde{\mathbf{x}}^{e}=\mathbf{L}_{m}\mathbf{d}.
\end{equation}
Here, ${\mathbf{L}}_{m}=2\mathbf{L}_{\theta}/(h_{1}+h_{4})$ denotes the corotational edge-based hinge curvature operator, where $\mathbf{L}_{\theta}$ is given by
\begin{equation} \label{eq10}
    \begin{bmatrix}
    \frac{1}{h_{1}}  & -\left( \frac{\lVert \tilde{\mathbf{X}}_{P3}\rVert}{\lVert \tilde{\mathbf{X}}_{23}\rVert h_{1}} + \frac{\lVert \tilde{\mathbf{X}}_{Q3}\rVert}{\lVert \tilde{\mathbf{X}}_{23}\rVert h_{4}} \right) &  -\left( \frac{\lVert \tilde{\mathbf{X}}_{P2}\rVert}{\lVert \tilde{\mathbf{X}}_{23}\rVert h_{1}} + \frac{\lVert \tilde{\mathbf{X}}_{Q2}\rVert}{\lVert \tilde{\mathbf{X}}_{23}\rVert h_{4}} \right) &  \frac{1}{h_{4}}  
    \end{bmatrix}
\end{equation}
and 
$\mathbf{N}=\mathbf{I}_{4\times 4} \otimes \mathbf{n}_{\tilde{z}}$. $\mathbf{I}_{4\times 4}$ is the fourth order identity matrix, $\otimes$ is the Kronecker product operator and the transverse displacement vector of the edge-based stencil $\mathbf{d}=\begin{bmatrix}\tilde{w}_{1}&\tilde{w}_{2}&\tilde{w}_{3}&\tilde{w}_{4}\end{bmatrix}^{T}$ measures the deviations of the nodes away from the corotational shell stencil in the current configuration (see Figure~\ref{fig:corotational_edge_based_hinge}). 

To conveniently get the derivatives of curvature, we express the transverse displacement vector using the world coordinates, leading to the directional curvature
\begin{equation} \label{eq11}
    \kappa_{{m}}=\mathbf{L}_{m}\mathbf{N}^{T}(\mathbf{x}^{e} - \mathbf{x}_{c}^{e}).
\end{equation}
In the geometry view, $\mathbf{x}^{e}_{c}$ lies in the tangential plane of the current corotational frame. Since $\mathbf{N}$ contains the normal directions of the current corotational frame, it's obvious that $\mathbf{N}^{T}\mathbf{x}^{e}_{c}=0$. We can obtain that
\begin{equation}  \label{eq12}
    \mathbf{L}_{m}\mathbf{N}^{T}\mathbf{x}^{e}_{c}=\mathbf{0}.
\end{equation} 
Thus, the curvature of the edge-based shell stencil in the current configuration is 
\begin{equation}  \label{eq13}
    \kappa_{{m}}=\mathbf{L}_{m}\mathbf{N}^{T}\mathbf{x}^{e}.
\end{equation}

Similarly, the curvature of the edge-based shell stencil in the initial configuration is 
\begin{equation}  \label{eq14}
    (\kappa_{{m}})_{0}=\mathbf{L}_{m}\mathbf{d}_{0}=\mathbf{L}_{m}\mathbf{N}_{0}^{T}\mathbf{X}^{e},
\end{equation}
where $\mathbf{N}_{0}=\mathbf{I}_{4\times 4} \otimes \mathbf{n}_{\tilde{Z}}$. In fact, $\mathbf{d}_{0}=\mathbf{N}_{0}^{T}\mathbf{X}^{e}$ measures the deviations of the edge-based shell stencil away from the corotational shell stencil in the initial configuration.

The curvature derivation process for the corotational edge-based hinge bending model can also be generalized to the corotational FVM/smoothed hinge bending model. The difference lies in quantifying curvature from the edge-based shell stencil to the triangle-centered shell stencil. For the corotational FVM/smoothed hinge bending model, the direction of the $\tilde{z}$-axis in the current corotational frame is exactly given by the normal of the central triangle $T_{123}$, i.e., $\mathbf{n}_c$ in Eq. (\ref{eq8}) should be replaced by
\begin{equation} \label{eq15}
    \mathbf{n}_c = \mathbf{x}_{12} \times \mathbf{x}_{13}.
\end{equation}

\noindent \textbf{Corotational edge-based hinge thin plate.}
The bending energy of the corotational edge-based hinge thin plate can be expressed as
\begin{equation} \label{eq16}
    \Psi_b^{EP} = \frac{1}{2} A_{\mathcal{E}} k_b \kappa_{{m}}^{2},
\end{equation}
where $A_{\mathcal{E}}$ is the total area of the edge stencil in the initial configuration. The bending rigidity is defined as $k_b = {Eh^{3}}/{[12(1-\nu^{2})]}$ with $E$ is the Young's modulus and $\nu$ is the Poisson's ratio. 
By applying Eq. (\ref{eq13}), the bending energy in Eq. (\ref{eq16}) can be discretized as
\begin{equation} \label{eq17}
     \Psi_b^{EP} = \frac{A_{\mathcal{E}}}{2} k_b\kappa_{{m}}^{2} = \frac{A_{\mathcal{E}}}{2}k_b (\mathbf{x}^{e})^{T}\mathbf{N}\mathbf{L}_{m}^{T} \mathbf{L}_{m} \mathbf{N}^{T} \mathbf{x}^{e}.
\end{equation}
This can be further regrouped as
\begin{equation} \label{eq18}
     \Psi_b^{EP} = \frac{A_{\mathcal{E}}}{2} k_b\kappa_{{m}}^{2} = \frac{A_{\mathcal{E}}}{2}k_b (\mathbf{x}^{e})^{T}(\mathbf{L}_{m}^{T}\mathbf{L}_{m}\otimes\mathbf{I}) \mathbf{N}\mathbf{N}^{T} \mathbf{x}^{e},
\end{equation}
where $\mathbf{I}$ is the 3rd identity matrix.

For any point within the shell stencil under the small strain/curvature assumption, the relation 
\begin{equation} \label{eq19}
    \mathbf{n}\mathbf{n}^{T}(\mathbf{x}-\mathbf{x}_{c})=\mathbf{n}\tilde{w} \simeq \mathbf{x}-\mathbf{x}_{c}
\end{equation}
holds, and with $\mathbf{L}_{m}\mathbf{N}^{T}\mathbf{x}^{e}_{c}=\mathbf{0}$ in Eq. (\ref{eq12}). Therefore, the bending energy can be simplified to
\begin{equation} \label{eq20}
    \Psi_b^{EP}=\frac{A_{\mathcal{E}}}{2} k_b (\mathbf{x}^{e})^{T}(\mathbf{L}_{m}^{T}\mathbf{L}_{m}\otimes\mathbf{I}) \mathbf{x}^{e}.
\end{equation}
This shows that the bending energy is quadratic in terms of the nodal positions, which implies that the Hessian of the bending energy is constant and given by
\begin{equation} \label{eq21}
    \frac{\partial^{2} \Psi_b^{EP}}{\partial \mathbf{x}^{e}{\partial (\mathbf{x}^{e})^{T}}} = k_b A_{\mathcal{E}}  \mathbf{L}_{m}^{T}\mathbf{L}_{m}\otimes\mathbf{I}.
\end{equation}

The gradient of the bending energy, being linear with respect to the nodal positions, is expressed as
\begin{equation} \label{eq22}
    \frac{\partial  \Psi_b^{EP}}{\partial \mathbf{x}^{e}}= (\frac{\partial^{2} \Psi_b^{EP}}{\partial \mathbf{x}^{e}{\partial (\mathbf{x}^{e})^{T}}}) \mathbf{x}^{e}.
\end{equation}

It is important to note that quantifying the edge-based curvature operator in the world frame leads to a discrete expression similar to that of the Quadratic Shell model~\cite{Bergou2006Aquadratic}. However, this formulation overestimates the bending energy by a factor of three, as numerical results will be discussed in detail in Section~\ref{Linear_plate_bending_benchmark}. The fundamental difference and accuracy discrepancy has been detailed in Appendix~\ref{appendixD}.

\noindent \textbf{Corotational edge-based hinge thin shell.}
The bending energy of the corotational edge-based hinge thin shell is
\begin{equation} \label{eq23}
    \Psi_b^{ES} = \frac{1}{2} A_{\mathcal{E}} k_b \epsilon_b^{2},
\end{equation}
where the curvature change $\epsilon_b$ is
\begin{equation} \label{eq24}
    \epsilon_b = {\kappa}_m-({\kappa}_{m})_{0}=\mathbf{L}_{m}\mathbf{N}^{T} {\mathbf{x}}^{e}-\mathbf{L}_{m}\mathbf{N}_{0}^{T} {\mathbf{X}}^{e}.
\end{equation}

The gradient of the bending energy is then
\begin{equation} \label{eq25}
    \frac{\partial \Psi_b^{ES}}{\partial \mathbf{x}^{e}}={A_{\mathcal{E}}} k_b  \frac{\partial \epsilon_b}{\partial \mathbf{x}^{e}}\epsilon_b,
\end{equation}
where the gradient of the curvature change is 
\begin{equation} \label{eq26}
    \frac{\partial \epsilon_b}{\partial \mathbf{x}^{e}} = ({\mathbf{x}}^{e})^{T}\frac{\partial \mathbf{N}}{\partial \mathbf{x}^{e}} \mathbf{L}_{m}^{T}+\mathbf{N}\mathbf{L}_{m}^{T}.
\end{equation}
Here, ${\partial \mathbf{N}}/{\partial \mathbf{x}^{e}} = \mathbf{I}_{4\times4}\otimes ({\partial \mathbf{n}_{\tilde{z}}}/{\partial \mathbf{x}^{e}})$ with the gradient of the normal ${\partial \mathbf{n}_{\tilde{z}}}/{\partial \mathbf{x}^{e}}$ is detailed in the Appendix~\ref{appendixB}.

Before deriving the Hessian of bending energy, substituting Eq. (\ref{eq24}) into Eq. (\ref{eq23}) yields 
\begin{equation} \label{eq27}
    \Psi_b^{ES} = \Psi_{flat}^{ES} + \Psi_{curved}^{ES},
\end{equation}
which decomposes into a flat part 
\begin{equation} \label{eq28}
    \Psi_{flat}^{ES} =\frac{A_{\mathcal{E}}}{2}k_b(\mathbf{x}^{e})^{T}(\mathbf{L}_{m}^{T}\mathbf{L}_{m}\otimes \mathbf{I})\mathbf{N}\mathbf{N}^{T}\mathbf{x}^{e}
\end{equation}
and a curved part $\Psi_{curved}^{ES}$ is
\begin{equation} \label{eq29}
    \begin{gathered}
    \frac{A_{\mathcal{E}}}{2}k_b(-2(\mathbf{x}^{e})^{T}(\mathbf{L}_{m}^{T}\mathbf{L}_{m}\otimes \mathbf{I})\mathbf{N}\mathbf{N}_{0}^{T}\mathbf{X}^{e}+(\mathbf{X}^{e})^{T}(\mathbf{L}_{m}^{T}\mathbf{L}_{m}^{}\otimes \mathbf{I})\mathbf{N}_{0}\mathbf{N}_{0}^{T}\mathbf{X}^{e}).
    \end{gathered}
\end{equation}

The flat part in Eq. (\ref{eq27}) is identical to the bending energy of the corotational edge-based hinge thin plate in Eq. (\ref{eq18}), similar note can be found in the Cubic Shell paper~\cite{Garg2007cubicshell}. Consequently, the simplified bending energy Hessian from the rest-flat version in Eq. (\ref{eq21}) can be used to perform a single bending energy Hessian assembly for the rest-curved version.

\noindent \textbf{Boundary conditions for the corotational edge-based hinge.}
In thin shell simulations, the most commonly used boundary conditions are the clamped boundary condition, the free boundary condition, and the simply supported boundary condition. The simply supported boundary condition can be effectively achieved by combining the free boundary condition with fixed boundary nodes, so we will focus on discussing the clamped and free boundary conditions. In the boundary edge stencil $MNLL'$, illustrated in Figure~\ref{fig:boundary_shell_patch}, the node $L'$ is a virtual node that is symmetrically positioned with respect to node $L$ across the midpoint $P$ of the boundary edge $MN$.

\textit{The Clamped Boundary Condition.} 
For a boundary edge stencil with a clamped boundary condition, a symmetric virtual transverse displacement is applied, corresponding to the virtual node $L'$ 
\begin{equation} \label{eq30}
    \tilde{w}_{L'} = \tilde{w}_{L}, 
\end{equation}
which ensures the preservation of the normal direction perpendicular to the boundary edge stencil. When this condition is applied to the boundary edge stencil, the corotational edge-based hinge curvature operator $\mathbf{L}_{m}$ is modified to
\begin{equation} \label{eq31}
    \mathbf{L}_{m}^{'}= \frac{1}{h_{1}}\begin{bmatrix}\frac{2}{h_{1}} & -\frac{\lVert \tilde{\mathbf{X}}_{P3} \rVert + \lVert \tilde{\mathbf{X}}_{Q3} \rVert}{\lVert \tilde{\mathbf{X}}_{23} \rVert h_{1}} &-\frac{\lVert \tilde{\mathbf{X}}_{P2} \rVert + \lVert \tilde{\mathbf{X}}_{Q2} \rVert}{\lVert \tilde{\mathbf{X}}_{23} \rVert h_{1}} &0 \end{bmatrix}.
\end{equation}
This zero-slope condition is also applicable in symmetric structural finite element analysis (FEA).

\textit{The Free Boundary Condition.}
For a free boundary edge, there is no bending energy in the boundary edge stencil, resulting in zero entries for both the gradient and Hessian related to this boundary (zero-curvature condition).

In conclusion, the modification of the curvature operator according to the missing nodes enables the condensation of boundary conditions at the stencil level. Boundary conditions are crucial in engineering shell simulations to achieve accurate results. However, their impact on the visual effects in computer animation is often negligible.

\subsection{Corotational FVM hinge bending model}

\begin{figure}
  \centering
  \includegraphics[width=.99\linewidth]{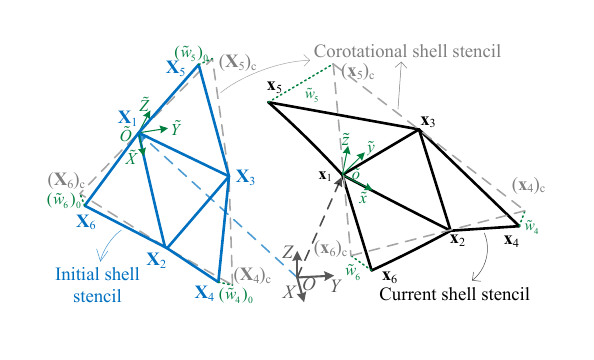}
  \caption{\label{fig:corotational_FVM_hinge}
           A triangle-centered shell stencil for modeling the corotational FVM hinge bending model deforms from its initial configuration (blue) to the current configuration (black). 
           $X-Y-Z$ is the world frame. For shell stencil in the current configuration, the $\tilde{x}-\tilde{y}-\tilde{z}$ frame (green) is the current corotational frame, where the $\tilde{x}$-axis aligns with the edge $\mathbf{x}_{12}$, and the $\tilde{z}$-axis direction is along the normal direction of the central triangle. The $\tilde{y}$-axis is determined by the right-hand rule. The corotational shell stencil (gray dashed line) is the projection of the shell stencil onto the $\tilde{x}-\tilde{y}$ plane. The point $(\mathbf{x}_4)_c$ represents the projection of $\mathbf{x}_4$ onto the $\tilde{x}-\tilde{y}$ plane, while the transverse displacement $\tilde{w}_{4}$ corresponds to the projection of the relative difference between $\mathbf{x}_4$ and $(\mathbf{x}_4)_c$ along the $\tilde{z}$-axis. Similar quantities for other points can be calculated in the same manner. Variables related to the shell stencil in the initial configuration can be extended from those in the current configuration.}
\end{figure}

\textit{Smooth Settings.} The curvature of the central triangle in the initial configuration can be evaluated using the finite volume method (FVM) within the corotational frame. The generalized curvature vector 
\begin{equation} \label{eq32}
    \boldsymbol{\kappa}_{0}=
    \frac{1}{A} \int_A \left(\begin{array}{c}
        \frac{\partial^2 \tilde{w}_{0}}{\partial \tilde{X}^2} \\ 
        \frac{\partial^2 \tilde{w}_{0}}{\partial \tilde{Y}^2} \\ 
        2 \frac{\partial^2 \tilde{w}_{0}}{\partial \tilde{X} \partial \tilde{Y}}
        \end{array}\right) dA
\end{equation}
is derived through variation the virtual work principle governing the bending deformation. Here $A$ is the area of the central triangle in the initial configuration. By applying the Stokes's theorem, Eq. (\ref{eq32}) can be rewritten as
\begin{equation} \label{eq33}
    \boldsymbol{\kappa}_{0}=\frac{1}{A} \int_{\Gamma} \mathbf{T} \binom{\frac{\partial \tilde{w}_{0}}{\partial \tilde{X}}}
    {\frac{\partial \tilde{w}_{0}}{\partial \tilde{Y}}}  d \Gamma,
\end{equation}
where $\Gamma$ is the boundary of the central triangle in the initial configuration, and $\mathbf{T}$ is defined as 
\begin{equation} \label{eq34}
    \mathbf{T}=\left[\begin{array}{cc}m_{s} & 0 \\0 & m_{t} \\m_{t} & m_{s} \end{array}\right],
\end{equation}
with the normalized normal $\tilde{\mathbf{m}}=\left[\begin{array}{cc}m_{s} & m_{t} \end{array}\right]^{T}$ outward to the boundary $\Gamma$ surrounding the central triangle in the $\tilde{X}-\tilde{Y}$ plane of the initial corotational frame. The ${s}$-axis is tangential to the boundary $\Gamma$ and ${t}$-axis is perpendicular to the ${s}$-axis.
The gradient of the transverse displacement can be transformed from the directional gradient, i.e.,
\begin{equation} \label{eq35}
    \binom{\frac{\partial \tilde{w}_{0}}{\partial \tilde{X}}}{\frac{\partial \tilde{w}_{0}}{\partial \tilde{Y}}}=
    \left[\begin{array}{cc}m_{{{s}}} & -m_{{t}} \\m_{{t}} & m_{{s}}\end{array}\right]
    \binom{\frac{\partial \tilde{w}_{0}}{\partial t}}{\frac{\partial \tilde{w}_{0}}{\partial s}},
\end{equation}
where ${\partial \tilde{w}_{0}}/{\partial t}$ and ${\partial \tilde{w}_{0}}/{\partial s}$ are respectively the rotation angle about the ${s}$-axis and ${t}$-axis.

\textit{Discrete Settings.} Given that there is no bending in each hinge edge (see Figure~\ref{fig:corotational_FVM_hinge}), i.e., ${\partial \tilde{w}}/{\partial t} \neq 0$ and ${\partial \tilde{w}}/{\partial s}=0$ in Eq. (\ref{eq35}), the generalized curvature of the central triangle in the initial configuration becomes
\begin{equation} \label{eq36}
    \boldsymbol{\kappa}_{0}= \sum_{i=1}^3 \left(\begin{array}{c}(m_{{s}})_{i}^2 \\ (m_{{t}})_{i}^2 \\2 (m_{{s}})_{i} (m_{{t}})_{i} \end{array}\right) 
    (\kappa_i)_0,
\end{equation}
where $(\kappa_i)_0=({\partial  \tilde{w}}/{\partial t})_{i}l_{i}/A$ is the directional curvature and $l_i $ is the length of the hinge edge in the initial configuration. It can be seen that the FVM approach superpositions the directional curvatures of the central triangle onto the generalized curvature in the corotational frame. To discrete the curvature and consider the adjacent triangles' contribution to the common edge, several discretized schemes can be found in~\cite{onate2000rotation, gingold2004discrete}.

In our approach, we first extend the Eq. (\ref{eq4}) to express the directional curvatures, $\boldsymbol{\kappa}_{0}$ can then be rewritten as
\begin{equation} \label{eq37}
    \boldsymbol{\kappa}_{0}=
    \mathbf{R}  (\boldsymbol{\kappa}_{p})_{0}=
    \mathbf{R} \left(\begin{array}{c}\frac{2(\theta_1)_{0}} {h_1+h_4}\\
    \frac{2(\theta_2)_{0}} {h_2+h_5}\\
    \frac{2(\theta_3)_{0}} {h_3+h_6}\end{array}\right),
\end{equation}
where $\boldsymbol{\kappa}_{p}$ denotes the directional curvature vector, $\theta_{i} \ (i=1,2,3)$ is the bend angle across each edge, $h_i \ (i = 1,2, \cdots ,6)$ is the triangle height in the initial configuration (see Figure~\ref{fig:geo_quan}), and the transform matrix $\mathbf{R}$ is
\begin{equation} \label{eq38}
    \mathbf{R} =\left[\begin{array}{ccc}(m_{{s}})^2_{1}& (m_{{s}})_{2}^2& (m_{{s}})_{3}^2 \\ (m_{{t}})_{1}^2 & (m_{{t}})_{2}^2 & (m_{{t}})_{3}^2 \\2 (m_{{s}})_{1} (m_{{t}})_{1}&2 (m_{{s}})_{2} (m_{{t}})_{2} & 2 (m_{{s}})_{3} (m_{{t}})_{3}\end{array}\right].
\end{equation}
Finally, under the small strain/curvature assumption, each directional curvature in Eq. (\ref{eq37}) is linearized using the corotational edge-based hinge approach described in Section~\ref{Corotational_edge-based_hinge_bending_model}.

\noindent \textbf{Corotational FVM hinge thin plate.} The bending energy of the corotational FVM hinge thin plate is
\begin{equation} \label{eq39}
    \Psi_\mathrm{b}^{FP}=\frac{A}{2}\boldsymbol{\kappa}_{p}^{T} \mathbf{D}_b^{p} \boldsymbol{\kappa}_{p}.
\end{equation}
Here $\mathbf{D}_b^{p}=\mathbf{R}^{T} \mathbf{D}_{\mathrm{b}}\mathbf{R}$, and the directional curvature vector in the current configuration can be discretized by
\begin{equation} \label{eq40}
    \boldsymbol{\kappa}_{p} = \mathbf{L}_{p}  \mathbf{N}^{T} \mathbf{x}^{e},
\end{equation}
where $\mathbf{N} =\mathbf{I}_{6\times 6} \otimes \mathbf{n}_{\tilde{z}}$ with $\mathbf{n}_{\tilde{z}}$ normalized as described $\mathbf{n}_c$ in Eq. (\ref{eq15}). $\mathbf{I}_{6\times 6}$ is a sixth order identity matrix. The entries of the directional curvature operator $\mathbf{L}_{p}$ are detailed in Appendix~\ref{appendixC}. Consequently, the generalized curvature in Eq. (\ref{eq36}) can be discretized by the corotational FVM hinge curvature operator $\mathbf{R}\mathbf{L}_{p}$.

In deriving the derivatives of the bending energy of corotational FVM hinge thin plate, we also obtain the constant bending energy Hessian corresponding to the corotational edge-based hinge thin plate in Section~\ref{Corotational_edge-based_hinge_bending_model}, i.e.,
\begin{equation} \label{eq41}
    \frac{\partial^{2} \Psi_b^{FP}}{\partial \mathbf{x^\mathrm{e}}{\partial (\mathbf{x}^{e})^{T}}}=A (\mathbf{L}_{p}^{T} \mathbf{D}_{\mathrm{b}}^{p} \mathbf{L}_{p} \otimes \mathbf{I}).
\end{equation}
The operation in Eq. (\ref{eq22}) can similarly be employed to compute the linear gradient of the corotational FVM hinge thin plate model.

\noindent \textbf{Corotational FVM hinge thin shell.}
The bending energy of the corotational FVM hinge thin shell is
\begin{equation} \label{eq42}
    \Psi_{{b}}^{FS}=\frac{A}{2} (\boldsymbol{\varepsilon}_{b}^{FS})^{T} \mathbf{D}_{\mathit{b}}^{} \boldsymbol{\varepsilon}_{b}^{FS}
\end{equation}
where the curvature change vector $\boldsymbol{\varepsilon}_b^{FS}$ is
\begin{equation} \label{eq43}
    \boldsymbol{\varepsilon}_b^{FS} = \mathbf{R}\boldsymbol{\kappa}_{p}-\mathbf{R}(\boldsymbol{\kappa}_{p})_{0}=\mathbf{R}\mathbf{L}_{p}\mathbf{N}^{T} {\mathbf{x}}^{e}-\mathbf{R}\mathbf{L}_{p}\mathbf{N}_{0}^{T} {\mathbf{X}}^{e}
\end{equation}
with $\mathbf{N}_{0}=\mathbf{I}_{6\times 6} \otimes \mathbf{n}_{\tilde{Z}}$.
The gradient of the bending energy of corotational FVM hinge thin shell is generalized from the corotational edge-based hinge thin shell, as follows
\begin{equation} \label{eq44}
    \frac{\partial \Psi_{{b}}^{FS}}{\partial \mathbf{x}^\mathrm{e}}=A\frac{\partial (\boldsymbol{\varepsilon}_b^{FS})^{T}}{\partial \mathbf{x}^{e}}\mathbf{D}_{{b}}^{}  \boldsymbol{\varepsilon}_b^{FS},
\end{equation}
where the gradient of the curvature change vector is
\begin{equation} \label{eq45}
    \frac{\partial \boldsymbol{\varepsilon}_b^{FS}}{\partial \mathbf{x}^{e}}=({\mathbf{x}}^{e})^{T}\frac{\partial \mathbf{N}}{\partial \mathbf{x}^{e}} \mathbf{L}_{p}^{T}\mathbf{R}^{T}+\mathbf{N}\mathbf{L}_{p}^{T}\mathbf{R}^{T}.
\end{equation}

The bending energy Hessian remains the same as in the corotational FVM hinge thin plate model, for reasons similar to those discussed in the corotational edge-based hinge bending model in Section~\ref{Corotational_edge-based_hinge_bending_model}.

\noindent \textbf{Boundary conditions for corotational FVM hinge.}
For boundary triangle-centered stencils where at least one triangle is missing along the boundary edge, the boundary conditions from the corotational edge-based hinge bending model can be applied to the boundary triangle stencil (see Figure~\ref{fig:boundary_shell_patch}). It is crucial to set the mixed second derivative of curvature to zero for the boundary shell stencil with free edge boundary condition to ensure accuracy in passing the "engineering shell obstacle benchmark" tests, shown in Section~\ref{Geometrically_Non-linear_Benchmarks}.

\subsection{Corotational smoothed hinge bending model}

\begin{figure}
  \centering
  \includegraphics[width=.99\linewidth]{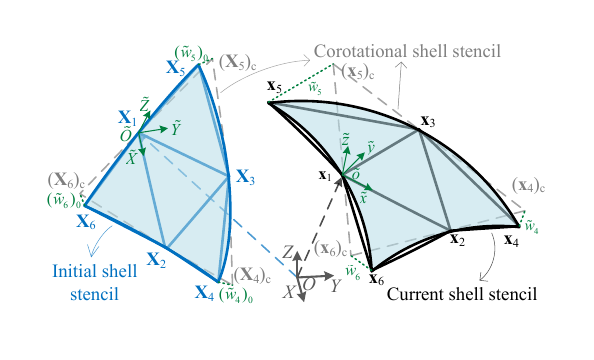}
  \caption{\label{fig:corotational_smoothed_hinge}
           A triangle-centered shell stencil, fitted to a quadratic surface (light blue), models the corotational smoothed hinge bending model as it deforms from its initial configuration (blue) to the current configuration (black).
           $X-Y-Z$ is the world frame, and $\tilde{x}-\tilde{y}-\tilde{z}$ (green) is the corotational frame for the shell stencil in the current configuration, capturing nodal deviations relative to the projected corotational shell stencil (gray dashed line). $(\mathbf{x}_4)_c$ is the projection of $\mathbf{x}_4$ onto the $\tilde{x}-\tilde{y}$ plane, while $\tilde{w}_{4}$ is the relative difference between $\mathbf{x}_4$ and $(\mathbf{x}_4)_c$ projected along the $\tilde{z}$-axis. Similar quantities for other points can be computed, and the shell stencil in the initial configuration variables can be extended from the current configuration.}
\end{figure}

In the smoothed hinge bending model, the generalized curvature vector 
\begin{equation} \label{eq46}
    \boldsymbol{\kappa}_{0} = \left(\begin{array}{c}
        \frac{\partial^2\tilde{w}_{0}}{\partial \tilde{X}^2} \\
        \frac{\partial^2\tilde{w}_{0}}{\partial \tilde{Y}^2} \\
        2 \frac{\partial^2\tilde{w}_{0}}{\partial \tilde{X} \partial \tilde{Y}}
        \end{array}\right) 
\end{equation}
is derived directly from the area integration in Eq. (\ref{eq32}), rather than transforming the area integration to line integration using integration by parts in Eq. (\ref{eq33}). The generalized curvature vector in Eq. (\ref{eq46}) can be discretized as 
\begin{equation} \label{eq47}
    \boldsymbol{\kappa}_{0} = \mathbf{L}_{g} \mathbf{d}_{0},
\end{equation}
where $\mathbf{d}_{0}=\begin{bmatrix}(\tilde{w}_{1})_{0}&(\tilde{w}_{2})_{0}&(\tilde{w}_{3})_{0}&(\tilde{w}_{4})_{0}&(\tilde{w}_{5})_{0}&(\tilde{w}_{6})_{0}\end{bmatrix}^{T}$ is the vector of transverse displacements in the initial configuration. In the smoothed hinge model, a quadratic fitting surface is used to smooth the triangle-centered stencil (see Figure~\ref{fig:corotational_smoothed_hinge}) and $\mathbf{L}_{g}$ is denoted as corotational smoothed hinge curvature operator. So the transverse displacement field on the shell surface can be interpolated as
\begin{equation} \label{eq48}
    (\tilde{w})_{0}=c_1+c_2 \tilde{X}+c_3 \tilde{Y}+c_4 \tilde{X}^2 / 2+c_5 \tilde{Y}^2 / 2+c_6 \tilde{X} \tilde{Y} / 2,
\end{equation}
where $c_i \ (i=1,\cdots,6)$ are constants. The first three coefficients correspond to rigid-body motion, while the last three correspond to curvature within the quadratic surface. By introducing Eq. (\ref{eq48}) into Eq. (\ref{eq46}) with Eq. (\ref{eq47}), the generalized curvature vector can be rewritten as $\boldsymbol{\kappa}_{0} = \mathbf{L}_{g} \mathbf{d}_{0} = \begin{bmatrix} c_4 & c_5 & c_6 \end{bmatrix}^{T}$. Its three components can be expressed by converting the three directional curvature constituent parts from the directional curvature vector $(\boldsymbol{\kappa}_{p})_{0}$. By recalling Eq. (\ref{eq9}), Eq. (\ref{eq13}) and Eq. (\ref{eq40}), $(\boldsymbol{\kappa}_{p})_{0}$ can be rewritten as $(\boldsymbol{\kappa}_{p})_{0}=\mathbf{L}_{p}\mathbf{d}_{0}$. Consequently,
\begin{equation} \label{eq49}
    (\boldsymbol{\kappa}_{p})_{0}=\mathbf{L}_{p}\mathbf{d}_{0}
    =\mathbf{L}_{p}\mathbf{C}_p \left(\begin{array}{c} c_4 \\ c_5 \\ c_6 \end{array}\right)
    =\mathbf{L}_{p}\mathbf{C}_p\mathbf{L}_{g}\mathbf{d}_{0},
\end{equation}
where $\mathbf{C}_p$ is portrayed as a matrix $\left[\begin{array}{cccccc} \tilde{X}_1^2 / 2 & \tilde{Y}_1^2 / 2 & \tilde{X}_1 \tilde{Y}_1 / 2 \\\vdots & \vdots  & \vdots  \\ \tilde{X}_6^2 / 2 & \tilde{Y}_6^2 / 2 & \tilde{X}_6 \tilde{Y}_6 / 2\end{array}\right]$. 
We then explicitly get the corotational smoothed hinge curvature operator
\begin{equation} \label{eq50}
    \mathbf{L}_{g}=(\mathbf{L}_{p}\mathbf{C}_p)^{-1}\mathbf{L}_{p}
\end{equation} 
to express the generalized curvature of the smoothed hinge bending model in the initial configuration as
\begin{equation} \label{eq51}
    \boldsymbol{\kappa}_{0} = \mathbf{L}_{g} \mathbf{N}_{0}^T\mathbf{X}^{e}.
\end{equation}

It should be mentioned that the constant curvature of the quadratic surface can also be computed by the least square method to solve a sixth order linear system for each curvature operator. In practice, we find that our presented method is more stable compared to this approach. Another explicit formulation is presented in a follow-up note by Reisman et al. \cite{reisman2007note}, which continues the work of Grinspun et al. \cite{Grinspun2006ComputingDS}, they highlight certain drawbacks of the curvature operator fitted on the quadratic surface. Notably, near-conic degenerate configurations can lead to numerical instabilities and challenges in boundary condition treatment. Thanks to our small strain/curvature assumption within the corotational framework,  the bending energy Hessian matrix can be assembled once using the initial geometric data. A good mesh initialization can effectively mitigate the numerical issue. In our numerous numerical exercises, even with no special initialization, we find that our model remains stable. More detailed results can be found in the Section~\ref{numerical_results}. Additionally, we provide a concise treatment of boundary conditions on our corotational smoothed hinge curvature operator, which will be elaborated upon later.

\noindent \textbf{Corotational smoothed hinge thin plate.}
The bending energy of the corotational smoothed hinge thin plate is
\begin{equation} \label{eq52}
    \Psi_{{b}}^{SP}=\frac{A}{2} \boldsymbol{\kappa}_{}^{T} \mathbf{D}_{{b}}  \boldsymbol{\kappa}_{}.
\end{equation}

The destination of the corotational smoothed hinge bending energy Hessian 
\begin{equation} \label{eq53}
    \frac{\partial \Psi_{{b}}^{SP}}{\partial \mathbf{x}^\mathrm{e}}=A (\mathbf{L}_{g}^{T} \mathbf{D}_{{b}} \mathbf{L}_{g} \otimes \mathbf{I})
\end{equation}
is also constant, and the linear gradient can also be computed by the approach in Eq.(\ref{eq22}).

\noindent \textbf{Corotational smoothed hinge thin shell.}
The bending energy of the corotational smoothed hinge thin shell is
\begin{equation} \label{eq54}
    \Psi_{\mathrm{b}}^{SS}=\frac{A}{2} (\boldsymbol{\varepsilon}_{b}^{SS})^{T} \mathbf{D}_{\mathit{b}}^{} \boldsymbol{\varepsilon}_{b}^{SS},
\end{equation}
where the curvature change vector $\boldsymbol{\varepsilon}_{b}^{SS}$ is
\begin{equation} \label{eq55}
   \boldsymbol{\varepsilon}_{b}^{SS} = \boldsymbol{\kappa}-\boldsymbol{\kappa}_{0}=\mathbf{L}_{g}\mathbf{N}^{T} {\mathbf{x}}^{e}-\mathbf{L}_{g}\mathbf{N}_{0}^{T} {\mathbf{X}}^{e}.
\end{equation}

The gradient of the bending energy of corotational smoothed hinge thin shell is
\begin{equation} \label{eq56}
    \frac{\partial \Psi_{\mathrm{b}}^{SS}}{\partial \mathbf{x}^\mathrm{e}}=A \frac{\partial (\boldsymbol{\varepsilon}_{b}^{SS})^{T}}{\partial \mathbf{x}^{e}} \mathbf{D}_{{b}}  \boldsymbol{\varepsilon}_{b}^{SS},
\end{equation}
where the gradient of the curvature change vector is
\begin{equation} \label{eq57}
    \frac{\partial (\boldsymbol{\varepsilon}_{b}^{SS})^{T}}{\partial \mathbf{x}^{e}}=({\mathbf{x}}^{e})^{T}\frac{\partial \mathbf{N}}{\partial \mathbf{x}^{e}} \mathbf{L}_{g}^{T}+\mathbf{N}\mathbf{L}_{g}^{T}.
\end{equation}

The bending energy Hessian is the same as that of the corotational smoothed hinge thin plate model. The rationale behind this similarity is consistent with the explanation provided for the corotational edge-based hinge thin shell model.

\begin{figure}
  \centering
  \includegraphics[width=.75\linewidth]{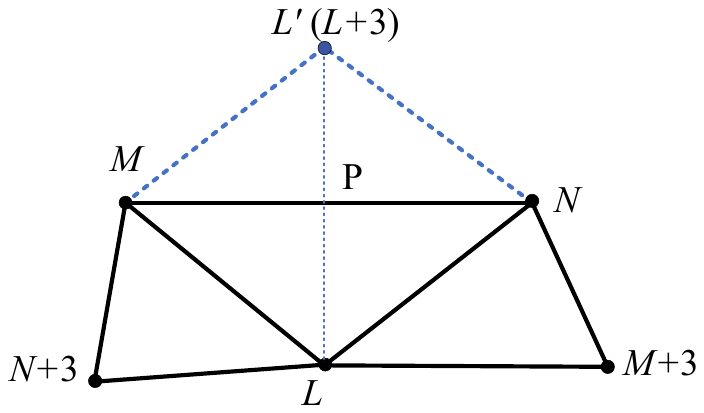}
  \caption{\label{fig:boundary_shell_patch}
           The artificial node \(L'\) (shown in blue) and node \(L\) are symmetric with respect to the midpoint \(P\) of the boundary edge \(NM\).
           }
\end{figure}
\noindent \textbf{Boundary conditions for corotational smoothed hinge.}
Within the context of the boundary edge $MN$ depicted in Figure~\ref{fig:boundary_shell_patch}, the mesh topology informs that the nodal label $L' = L + 3$. Consequently, the curvature in the boundary shell stencil can be expressed as 
\begin{equation} \label{eq58}
    \boldsymbol{\kappa}=\sum_{}\mathbf{L}_{gj}\tilde{w}_j+\mathbf{L}_{g(L+3)}\tilde{w}_{L+3}, \ \text{for}\  j=N,M,L,N+3,M+3.
\end{equation}
Here, $\mathbf{L}_{gj}$ is the $j$ column of the curvature operator matrix $\mathbf{L}_{g}$.

\textit{The Clamped Boundary Condition.} To integrate the zero-slope condition, using Eq. (\ref{eq30}), the modified curvature with a clamped edge can be written as
\begin{equation} \label{eq59}
\begin{split}
    \boldsymbol{\kappa}_{clamp}=&(\mathbf{L}_{gL}+\mathbf{L}_{g(L+3)})\tilde{w}_{L}+\mathbf{L}_{gM}\tilde{w}_{M}\\
                        &+\mathbf{L}_{gN}\tilde{w}_{N}+\mathbf{L}_{g(N+3)}\tilde{w}_{N+3}+\mathbf{L}_{g(M+3)}\tilde{w}_{M+3}.
\end{split}
\end{equation}

\textit{The Free Boundary Condition.} For the zero-curvature condition, the virtual transverse displacement of the missing node should satisfy
\begin{equation} \label{eq60}
    \tilde{w}_{L'} = 2\tilde{w}_{P} - \tilde{w}_{L},
\end{equation}
where $\tilde{w}_{P} = (\tilde{w}_{M}+\tilde{w}_{N})/2$ is the transverse displacement of the middle point $P$ of the boundary edge $NM$. We can deduce the modified curvature with a free edge as follows
\begin{equation} \label{eq61}
\begin{split}
    \boldsymbol{\kappa}_{free}=&(\mathbf{L}_{gN}+\mathbf{L}_{g(L+3)})\tilde{w}_{N}+(\mathbf{L}_{gM}+\mathbf{L}_{g(L+3)})\tilde{w}_{M}\\ 
                        &+(\mathbf{L}_{gL}-\mathbf{L}_{g(L+3)})\tilde{w}_{L}+\mathbf{L}_{g(N+3)}\tilde{w}_{N+3}+\mathbf{L}_{g(M+3)}\tilde{w}_{M+3}.
\end{split}
\end{equation}

If a boundary triangle includes one more boundary edge, the operations outlined in Eq. (\ref{eq59}) and Eq. (\ref{eq61}) can be superimposed.

\begin{figure*} [htbp]
  \centering
  \includegraphics[width=.99\linewidth]{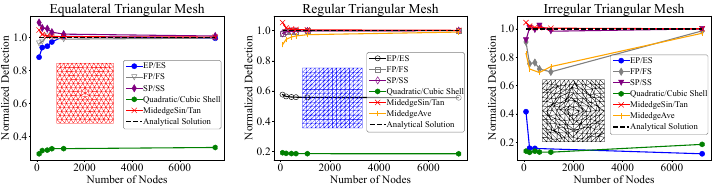}
  \caption{\label{fig:linear_plate_bending}
           \textit{Linear plate bending benchmark.} Convergence and consistency analysis of a simply supported linear plate under uniform load across three different mesh structures. The vertical axis represents the computed deflection normalized by the analytical value.
           EP/ES yield identical predictions, as do FP/FS and SP/SS. MidedgeSin and MidedgeTan also produce the same predictions, so a single marker is used for each group. MidedgeAve results in "NaN" values on the equilateral triangular mesh, and therefore its corresponding line is omitted.}
\end{figure*}

\section{Implementations}
\textit{Dynamics Simulation.} The incremental potential (IP)~\cite{kane2000variational} for elastodynamic simulations can be expressed as
\begin{equation} \label{eq62}
    E(\boldsymbol{x}) = \frac{1}{2}(\boldsymbol{x}-\hat{\boldsymbol{x}})^{T} \boldsymbol{M}(\boldsymbol{x}-\hat{\boldsymbol{x}}) + \Delta t^2 \Psi_{{elastic}} + B(\boldsymbol{x}) + D(\boldsymbol{x}),
\end{equation}
where \(\boldsymbol{M} \in {R}^{3n \times 3n}\) is the mass matrix, \(\Delta t\) is the time step, \(\hat{\boldsymbol{x}} = \boldsymbol{x}^t + \Delta t \boldsymbol{v}^t + \Delta t^2 \boldsymbol{M}^{-1} \boldsymbol{f}_{{ext}}\) represents the predicted position from the implicit Euler integration, with \(\boldsymbol{f}_{{ext}} \in {R}^{3n}\) is the external force, $\boldsymbol{x}^t \in {R}^{3n}$ and $\boldsymbol{v}^t \in {R}^{3n}$ stack the nodal positions and nodal velocities at time $t$, respectively. \(\Psi_{{elastic}}\) refers to elastic potential, which contains the elastic shell energy $\Psi_{shell}$. \(B(\boldsymbol{x})\) and \(D(\boldsymbol{x})\) are respectively the contact barrier potential and the friction potential~\cite{Li2020IPC, Li2021CIPC}.
In this unified formulation, both elastic and contact interactions are incorporated. The nodal positions at time $t+1$ are updated by minimizing the total potential
\begin{equation} \label{eq63}
    \boldsymbol{x}^{t+1} = \operatorname{argmin}_{\boldsymbol{x}} E(\boldsymbol{x}),
\end{equation}
where the solution is obtained iteratively using a Newton-type solver along with a continuous collision detection filter, ensuring intersection-free trajectories.

\textit{Linear and Quasi-static Simulations.} The quasi-static simulation is used to evaluate the accuracy and efficiency of different formulations in Section~\ref{Quantitative_Experiments} in which the contact is not involved, and equilibria are determined by setting the gradient of the total potential to zero, i.e.,
\begin{equation} \label{eq64}
    \frac{\partial \Psi_{{shell}}}{\partial \boldsymbol{x}} + \boldsymbol{f}_{ext}= \boldsymbol{0}.
\end{equation}
As quantitative benchmark problems will be compared with the general FEA package~\textcopyright {ABAQUS} in the engineering field, we use a standard Newton-Raphson method~\cite{bergan1978solution} under one load step to solve this system. The convergence criteria is $\lVert \partial \Psi_{\text{shell}} / \partial \boldsymbol{x} + \boldsymbol{f}_{\text{ext}} \rVert < \epsilon_{f} $, where $\epsilon_f$ is the residual force tolerance.
At each Newton iteration, $\Delta \boldsymbol{x}$ is the incremental displacement. If its infinity norm $\lVert \Delta \boldsymbol{x} \rVert_{\infty}$ exceeds the incremental displacement limit $\epsilon_u$, the line search step is scaled by $\epsilon_u / \lVert \Delta \boldsymbol{x} \rVert_{\infty}$. When the solution is far from equilibrium, this method can effectively reduce the step size for geometrically non-linear problems, including bending-dominated problems.
For the linear plate bending test, the solution can be obtained with one linear system solve. 

\section{Numerical Experiments} \label{numerical_results}

Quantitative benchmark problems are compared with reference solutions, including the analytical solution for the linear plate bending benchmark and results from \textcopyright{ABAQUS} for geometrically non-linear benchmarks. We also compare against state-of-the-art formulations from the discrete geometry shell library, libshell~\cite{Chen2018Physical}, which includes three types of formulations: MidedgeTan, MidedgeSin, and MidedgeAve. Additionally, we consider Quadratic and Cubic Shells~\cite{Bergou2006Aquadratic, Garg2007cubicshell}, all of which employ constant bending energy Hessians. These comparisons demonstrate the accuracy and efficiency of our models. Furthermore, qualitative numerical experiments highlight the stability and robustness of our models in elastodynamics simulations. All formulations are implemented on the codebase of libshell~\cite{Chen2018Physical, Chen2021WTF} for quantitative comparison, and all our formulations are integrated into the C-IPC~\cite{Li2021CIPC} for qualitative experiments. All experiments were performed on a workstation equipped with an AMD Ryzen Threadripper 3970X CPU (2.2 GHz, 32 cores) and 128 GB of RAM. 

\subsection{Quantitative Experiments} \label{Quantitative_Experiments}

\noindent \textbf{Linear plate bending benchmark.} \label{Linear_plate_bending_benchmark}
We investigate the analytical linear benchmark using three distinct mesh structures (refer to Figure~\ref{fig:linear_plate_bending}) to evaluate how our models and existing formulations depend on the mesh structure and to observe their convergence behaviour under mesh refinement. The membrane deformation can be neglected in this pure bending test, so the membrane formulation is excluded. Since comparable studies do not provide the detailed implementation of clamped boundary conditions, we apply simply supported  
boundary condition on the entire boundary of the square plate, which is subjected to a uniform load perpendicular to its plane, to ensure a fair comparison. The square plate has an edge length of $a=8$, with a uniform load of $B=9.81$ acting on the body. The material properties are defined by $E=2 \times 10^{11}$, $\nu=0.3$, and $h=0.01$. The analytical solution for the maximum deflection is $0.048744Ba^4(1-\nu^2)/(Eh^3)$~\cite{timoshenko1959theory}.

As shown in Figure~\ref{fig:linear_plate_bending}, both the EP and ES models pass the test only on the equilateral triangular mesh. However, the Quadratic Shell and Cubic Shell models fail all tests, even on the equilateral triangular mesh. The primary issue, as mentioned in Appendix~\ref{appendixD}, is that the bending energy formulation in the Quadratic and Cubic Shell models is three times higher than ours. The MidedgeAve model also fails on the equilateral triangular mesh due to numerical issues resulting in "NaN" values. On the regular triangular mesh, our FP, FS, SP, and SS models perform slightly better than the MidedgeSin and MidedgeTan formulations. On the irregular mesh, MidedgeSin and MidedgeTan exhibit better consistency than other methods. Among our models, SP and SS outperform FP and FS, which have comparable performance to the MidedgeAve formulation.

The data from the last column in Table~\ref{tab:combined_table} highlights that our FP, FS, SP, and SS models are nearly twice as fast as the MidedgeSin and MidedgeTan formulations in one linear system solve on the equilateral triangular mesh. Additionally, the EP/ES and Quadratic/Cubic Shell models in the edge-based stencil demonstrate exceptional speed. The numerical performance differences across various mesh tessellations primarily arise from the curvature operators used in these formulations.

\noindent \textbf{Geometrically non-linear benchmarks.} \label{Geometrically_Non-linear_Benchmarks}
In this subsection, we aim to verify the expected accuracy and efficiency of our models in geometrically non-linear cases (see Table~\ref{tab:combined_table}). The tested cases are derived from engineering obstacles~\cite{sze2004popular}. The reference solutions are obtained using the S4R shell element in \textcopyright{ABAQUS} with a sufficiently high mesh density. The underlined geometry of S4R is quadrilateral, so the tested mesh is generated by splitting each quadrilateral into two triangles. For each simulation, the residual force tolerance $\epsilon_f$ is $0.001$, and the incremental displacement limit is $\epsilon_u=0.1$.

\begin{table*} [htbp]
\centering
\begin{tabular}{llcl|ccc|cl}
\toprule
\multirow{2}{*}{\textbf{Model}} & \multicolumn{2}{c}{\textbf{Cantilever}} & \multicolumn{1}{c}{} & \multicolumn{3}{c}{\textbf{Hemisphere Shell}} & \multicolumn{1}{c}{\textbf{Linear Plate Bending}} \\
\cmidrule(lr){2-3} \cmidrule(lr){5-7} \cmidrule(lr){8-8}
& $w_{tip}$  & Iterations  (Time) & & $u_{min}$ & $v_{max}$ & Iterations (Time) & Time (7459 nodes) \\
\midrule
EP              & 5.388  & 61 (0.293s)    &  &           &          &               & 1.758s \\
ES              & 5.387  & 61 (0.343s)    &  & -4.072    & 2.799    & 56 (27.334s)  & 1.778s \\
FP              & 6.056  & 68 (0.412s)    &  &           &          &               & 3.713s \\
FS              & 6.072  & 68 (0.464s)    &  & -5.752    & 3.403    & 84 (44.153s)  & 3.788s \\
SP              & 6.055  & 67 (0.535s)    &  &           &          &               & 3.620s \\
SS              & 6.072  & 67 (0.595s)    &  & -5.923    & 3.534    & 87 (45.035s)  & 3.759s \\
\midrule
Quadratic Shell & 2.510  & 29 (0.145s)    &  &           &          &               & 1.757s \\
Cubic Shell     & 2.510  & 29 (0.153s)    &  & -3.193    & 2.414    & 51 (17.784s)  & 1.761s \\
MidedgeTan      & 5.405  & 77 (2.545s)    &  & -5.831    & 3.422    & 93 (173.531s) & 6.963s \\
MidedgeSin      & 5.418  & 77 (2.596s)    &  & -5.886    & 3.451    & 94 (180.139s) & 6.979s \\
MidedgeAve      & 5.380  & 76 (2.213s)    &  & -5.564    & 3.331    & 93 (105.898s) & NaN    \\
\midrule
\textit{\textcopyright{ABAQUS} S4R}     & \textit{6.012} & \textit{106} & & \textit{-5.902} & \textit{3.406} & \textit{112} & \\
\bottomrule
\end{tabular}
\caption{Displacement, Newton iteration and time data for the Cantilever under End Shear Force, Hemispherical Shell under Alternating Radial Forces, and Linear Plate Bending examples. "NaN" represents numerical issues encountered by the model. The results of \textit{\textcopyright{ABAQUS} S4R} act as reference solutions. $w_{\text{tip}}$ represents the displacement along the positive Z-direction of the midpoint on the right-hand side of the cantilever plate. $u_{\min}$ is the maximum displacement along the negative X-direction and $v_{\max}$ is the maximum displacement along the positive Y-direction.}
\label{tab:combined_table}
\end{table*}

\begin{figure} 
  \centering
  \includegraphics[width=.99\linewidth]{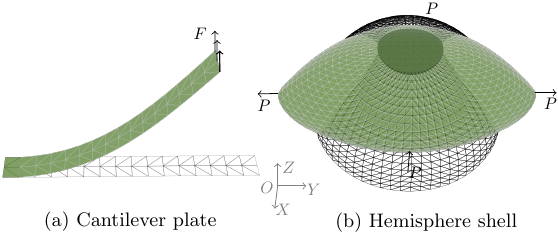}
  \caption{\label{fig:geometricall_non_linear_benchmarks}
           \textit{Geometrically non-linear benchmarks.} (a) Cantilever plate (51 nodes) subjected to an end shear force $F$ and (b) hemisphere shell with a $18^{\circ}$ cut (1088 nodes) subjected to alternating radial forces $P$ are tested to evaluate the accuracy and efficiency of different bending formulations in geometrically non-linear analysis~\cite{sze2004popular}. The green structure with a white wireframe represents the deformed configuration, while the black wireframe illustrates the undeformed configuration for comparison. $X-Y-Z$ is the world frame.}
\end{figure}

\textit{Cantilever Subjected to End Shear Force.}
In this test, a flat plate of dimensions $10 \times 1$ is subjected to an end shear force, applied as concentrated loads of equal magnitude $F=4/3$ distributed across the nodes on the right side, as illustrated in Figure~\ref{fig:geometricall_non_linear_benchmarks}. The concentrated loads are along the $Z$-axis. The geometry is discretized into 51 nodes, with two adjacent rows at the left end clamped to enforce the hard constraints.
The material parameters are $E = 1.2 \times 10^6$, $\nu = 0.1$, and $h = 0.1$. As summarized in Table~\ref{tab:combined_table}, our FP, FS, SP, and SS models demonstrate superior accuracy compared to others. The MidedgeTan and MidedgeSin formulations, which offer the second-highest accuracy, are nearly five times slower than our FP, FS, SP, and SS models, thanks to their constant bending energy Hessians. Our EP and ES models rank third in accuracy, outperforming the MidedgeAve model. While the Quadratic and Cubic Shell models are very fast, they produce smaller deflections due to the overestimation of bending rigidity. If we scale down the bending rigidity of the Quadratic and Cubic Shell models, they can predict deflections comparable to those of our EP model. Among our models, EP, FP and SP, specifically designed for rest-flat shell configurations, perform slightly better in terms of accuracy and speed compared to ES, FS and SS, which can handle both rest-flat and rest-curved shells. 

\textit{Hemispherical Shell Subjected to Alternating Radial Forces.}
To test the performance of the rest-curved shell models, we simulate a hemispherical shell with radius $R = 10$ and an $18^{\circ}$ circular cutout at the pole. The shell is subjected to alternating radial point forces of $P = 200$ at $90^{\circ}$ intervals (as shown in Figure~\ref{fig:geometricall_non_linear_benchmarks}). Two point forces along the $X$-axis induce compression, while two along the $Y$-axis induce tension. To minimize boundary condition effects across different formulations, the entire shell structure is analyzed instead of only a quarter section. Boundary conditions are applied as follows: for nodes lying in the $Y-Z$ plane, the $X$-direction DoFs are fixed; for nodes in the $X-Z$ plane, the $Y$-direction DoFs are fixed. Additionally, for nodes on the top circular cut that lie in the $Y-Z$ plane, the $Z$-direction DoFs are constrained to ensure equivalence with the benchmark case provided in~\cite{sze2004popular}. The shell geometry is discretized using 1088 nodes. The material parameters are $E = 6.825 \times 10^7$, $\nu = 0.3$, and $h = 0.04$. As shown in Table~\ref{tab:combined_table}, our ES model outperforms the Cubic Shell models in terms of accuracy. However, the ES model still deviates more from the reference solution. While the MidedgeTan and MidedgeSin models provide more accurate results overall, our FS and SS models deliver competitive accuracy with nearly four times the computational speed of the MidedgeTan and MidedgeSin models.

\subsection{Qualitative Experiments}

We introduce two challenging codimensional simulation benchmarks from C-IPC~\cite{Li2021CIPC} to demonstrate the robustness and stability of our models. The first test case involves a flat geometry, which is well-suited to all of our models, while the second features a cylindrical geometry, requiring compatible models for rest-curved meshes. Both cases use material properties corresponding to cotton from C-IPC, with $E = 0.8 \text{MPa}$, $\nu = 0.243$, and a cloth density of $472.6 \, \text{kg/m}^3$. The simulations are time-stepped at $\Delta t =0.04\text{s}$, also consistent with the solver settings from the C-IPC study. Simulation videos can be found in the supplementary materials.

\begin{figure*}[htbp]
  \centering
  \includegraphics[width=.9\linewidth]{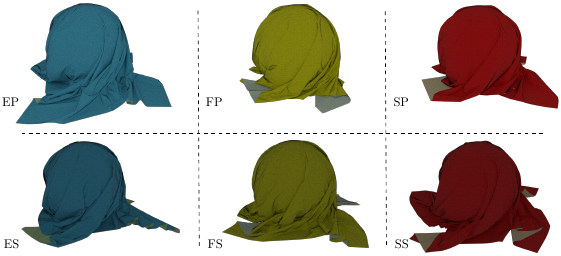}
  \caption{\label{fig:cloth_on_rotating_sphere}
           \textit{Cloth on rotating sphere and floor.} The images (the 100th frame) show the cloth’s response to being dropped onto a rotating sphere and floor (both with a friction coefficient $\mu = 0.4$). The cloth, with 85K nodes and a strain limit of 1.0608, is pulled inward by friction, generating a complex structure of wrinkles and folds. The top row (lighter) illustrates the cloth behaviour for corotational edge-based hinge thin plate (EP), corotational FVM hinge thin plate (FP), and corotational smoothed hinge thin plate (SP) models, while the bottom row displays results for corotational edge-based hinge thin shell (ES), corotational FVM hinge thin shell (FS), and corotational smoothed hinge thin shell (SS) models, highlighting the stability of each formulation in challenging cloth benchmark.
           }
\end{figure*}

\textit{Cloth on Rotating Sphere.}  
This test evaluates the robustness of our formulations (EP, ES, FP, FS, SP and SS) under extreme stress-test conditions, such as tight wrinkling, friction, and contact processing, following the setup used in prior research~\cite{Bridson2002Robust, Li2021CIPC}. A square cloth with 85K nodes (strain limiting up to 1.0608) is dropped onto a sphere and floor, both having a friction coefficient $\mu = 0.4$. As the sphere rotates, friction draws the cloth inward, creating a complex structure of wrinkles and folds, effectively capturing fine details of the cloth's behaviour (see Figure~\ref{fig:cloth_on_rotating_sphere}).

\begin{figure*}
  \centering
  \includegraphics[width=.9\linewidth]{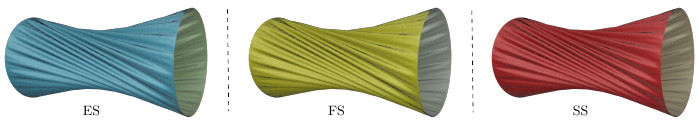}
  \caption{\label{fig:twisted_cylinder}
           \textit{Twisted cylinder.} Simulation (the 10th frame) of a 1m-wide, 0.25m-radius cotton cylinder (88K nodes) with a 1.5mm thickness offset. The cylinder is twisted at $72^\circ/s$ while the ends are drawn together at 5mm/s. Contact barrier is triggered at a 1mm threshold, and gravity is excluded to prevent sagging, resulting in pronounced wrinkling and folding, demonstrating our model’s robustness with rest-curved geometry. From left to right, the simulated frames are respectively generated by our corotational edge-based hinge thin shell (ES), corotational FVM hinge thin shell (FS), and corotational smoothed hinge thin shell (SS) models.
           }
\end{figure*}

\textit{Twisted Cylinder.}  
In this test, we simulate a cotton cylinder (1m width and 0.25m radius) with 88K nodes. The thickness offset is set to 1.5mm to account for geometric thickness, and the IPC~\cite{Li2020IPC} contact force is began at a threshold distance of 1mm. The cylinder is simultaneously twisted at a rate of $72^\circ/s$ while the two sides are brought together at 5mm/s. Gravity is excluded from the simulation to prevent sagging. As illustrated frame in Figure~\ref{fig:twisted_cylinder}, global wrinkling and folding effects emerge as the cylinder is deformed, showcasing the ability of our models (ES, FS and SS) to handle rest-curved geometry robustly. It's worthy to mention that SS generate 20 wrinkles, but ES and FS both give 19 waves.

\section{Conclusions}

In this study, our edge-stenciled models (EP and ES) are more accurate compared to the Quadratic and Cubic Shell models. 
The formulations of quadratic thin plate/shell (QTP/QTS), a variation of our EP/ES that quantifies the curvature operator in the world frame, are provided in Appendix~\ref{appendixD}. By introducing the formulations of QTP/QTS, the accuracy discrepancy between our edge-stenciled models with the Quadratic/Cubic Shells~\cite{Bergou2006Aquadratic, Garg2007cubicshell} is clarified.
Like the Quadratic and Cubic Shell models, the EP and ES models are computationally efficient. However, they share the same limitations common to all edge-based hinge approaches, such as sensitivity to mesh structure, as also discussed by Grinspun et al.~\cite{Grinspun2006ComputingDS}. Our triangle-stenciled models (FP, FS, SP, and SS) partially address this issue. Among these models, the consistency of FP and FS across different mesh patterns is weaker compared to SP and SS, which benefit from the smoothing effect of quadratic interpolation functions. 
Nevertheless, these models are constrained by small strain and small curvature assumptions, and failure modes may occur when the bend angle between a flap triangle and the central triangle exceeds $90^{\circ}$, leading to underestimation of the bending energy. Methods such as adaptive mesh refinement~\cite{Grinspun2002CHARMS, Narain2012arcsim, Ferguson2023} can be employed in regions of folding turns to mitigate this issue with lower mesh density. Despite these assumptions, the use of a corotational approach to handle large rotations offers significant advantages, allowing the bending energy Hessian to remain constant in Newton-type implicit solvers. 
Our experience indicates that the global constant bending energy Hessian assembled from those of stencils is more robust, helping to avoid potential numerical issues and achieve better accuracy~\cite{Grinspun2006ComputingDS, reisman2007note}, if a quality mesh is employed.
In our quasi-static simulations, we employ a basic Newton solver to evaluate the quantitative performance of different bending formulations in comparison to the shell element provided in \textcopyright{ABAQUS}. Therefore, we recommend integrating these formulations into a robust, well-designed solver to fully exploit their efficiency in practical applications.

Due to our models' simplicity, accuracy, efficiency, and generality, we anticipate that our models will have practical applications in both computer animation and specialized engineering simulations.

\section*{Acknowledgments}
The author sincerely thanks the anonymous reviewers for their valuable comments. Special gratitude is extended to Prof. K.Y. Sze for his insightful suggestions. The author also gratefully acknowledges the support provided by the Centre for Transformative Garment Production (TranGP).

\bibliographystyle{eg-alpha-doi} 
\bibliography{egbibsample}                 

\newcommand{\etalchar}[1]{$^{#1}$}
\begin{thebibliography}{\uppercase{BWH{\etalchar{*}}06b}}

\bibitem[BBHH11]{benson2011large}
\textsc{Benson D.~J., Bazilevs Y., Hsu M.-C., Hughes T.}:
\newblock A large deformation, rotation-free, isogeometric shell.
\newblock \emph{Computer Methods in Applied Mechanics and Engineering 200}, 13-16 (Mar 2011), 1367--1378.
\newblock \href {https://doi.org/10.1016/j.cma.2010.12.003} {\path{doi:10.1016/j.cma.2010.12.003}}.

\bibitem[BFA02]{Bridson2002Robust}
\textsc{Bridson R., Fedkiw R., Anderson J.}:
\newblock Robust treatment of collisions, contact and friction for cloth animation.
\newblock \emph{ACM TOG 21}, 3 (Jul 2002), 594–603.
\newblock \href {https://doi.org/10.1145/566654.566623} {\path{doi:10.1145/566654.566623}}.

\bibitem[BHBS78]{bergan1978solution}
\textsc{Bergan P., Horrigmoe G., Br{\aa}keland B., S{\o}reide T.}:
\newblock Solution techniques for non-linear finite element problems.
\newblock \emph{International Journal for Numerical Methods in Engineering 12}, 11 (1978), 1677--1696.
\newblock \href {https://doi.org/10.1002/nme.1620121106} {\path{doi:10.1002/nme.1620121106}}.

\bibitem[BHW94]{Breen1994Predicting}
\textsc{Breen D.~E., House D.~H., Wozny M.~J.}:
\newblock Predicting the drape of woven cloth using interacting particles.
\newblock In \emph{Proc. SIGGRAPH '94} (1994), p.~365–372.
\newblock \href {https://doi.org/10.1145/192161.192259} {\path{doi:10.1145/192161.192259}}.

\bibitem[BMF03]{Bridson2003Cloth}
\textsc{Bridson R., Marino S., Fedkiw R.}:
\newblock Simulation of clothing with folds and wrinkles.
\newblock In \emph{Proceedings of the 2003 ACM SIGGRAPH/Eurographics Symposium on Computer Animation} (2003), SCA '03, p.~28–36.
\newblock \href {https://doi.org/10.5555/846276.846281} {\path{doi:10.5555/846276.846281}}.

\bibitem[BW98]{BW98}
\textsc{Baraff D., Witkin A.}:
\newblock Large steps in cloth simulation.
\newblock In \emph{Proc. SIGGRAPH '98} (1998), p.~43–54.
\newblock \href {https://doi.org/10.1145/280814.280821} {\path{doi:10.1145/280814.280821}}.

\bibitem[BWH{\etalchar{*}}06a]{Bergou2006SigCourse}
\textsc{Bergou M., Wardetzky M., Harmon D., Zorin D., Grinspun E.}:
\newblock Discrete quadratic curvature energies.
\newblock In \emph{ACM SIGGRAPH 2006 Courses} (2006), SIGGRAPH '06, p.~20–29.
\newblock \href {https://doi.org/10.1145/1185657.1185663} {\path{doi:10.1145/1185657.1185663}}.

\bibitem[BWH{\etalchar{*}}06b]{Bergou2006Aquadratic}
\textsc{Bergou M., Wardetzky M., Harmon D., Zorin D., Grinspun E.}:
\newblock A quadratic bending model for inextensible surfaces.
\newblock In \emph{Proceedings of the Fourth Eurographics Symposium on Geometry Processing} (2006), SGP '06, p.~227–230.
\newblock \href {https://doi.org//10.5555/1281957.1281987} {\path{doi:/10.5555/1281957.1281987}}.

\bibitem[CCK{\etalchar{*}}21]{Chen2021WTF}
\textsc{Chen Z., Chen H.-Y., Kaufman D.~M., Skouras M., Vouga E.}:
\newblock Fine wrinkling on coarsely meshed thin shells.
\newblock \emph{ACM TOG 40}, 5 (Aug 2021).
\newblock \href {https://doi.org/10.1145/3462758} {\path{doi:10.1145/3462758}}.

\bibitem[CK02]{Choi2002Stable}
\textsc{Choi K.-J., Ko H.-S.}:
\newblock Stable but responsive cloth.
\newblock \emph{ACM TOG 21}, 3 (Jul 2002), 1--19.
\newblock \href {https://doi.org/10.1145/566654.566624} {\path{doi:10.1145/566654.566624}}.

\bibitem[COS00]{cirak2000subdivision}
\textsc{Cirak F., Ortiz M., Schr{\"o}der P.}:
\newblock Subdivision surfaces: a new paradigm for thin-shell finite-element analysis.
\newblock \emph{International Journal for Numerical Methods in Engineering 47}, 12 (Apr 2000), 2039--2072.
\newblock \href {https://doi.org/10.1002/(SICI)1097-0207(20000430)47:12<2039::AID-NME872>3.0.CO;2-1} {\path{doi:10.1002/(SICI)1097-0207(20000430)47:12<2039::AID-NME872>3.0.CO;2-1}}.

\bibitem[Cri97]{CrisfieldFEM1997}
\textsc{Crisfield M.~A.}:
\newblock \emph{Non-Linear Finite Element Analysis of Solids and Structures: Advanced Topics}.
\newblock John Wiley \& Sons, Inc., 1997.

\bibitem[CSvRV18]{Chen2018Physical}
\textsc{Chen H.-Y., Sastry A., van Rees W.~M., Vouga E.}:
\newblock Physical simulation of environmentally induced thin shell deformation.
\newblock \emph{ACM TOG 37}, 4 (jul 2018).
\newblock \href {https://doi.org/10.1145/3197517.3201395} {\path{doi:10.1145/3197517.3201395}}.

\bibitem[EKS03]{Etzmu2003fast}
\textsc{Etzmu{\ss} O., Keckeisen M., Stra{\ss}er W.}:
\newblock A fast finite element solution for cloth modelling.
\newblock In \emph{Proceedings of the 11th Pacific Conference on Computer Graphics and Applications} (2003), PG '03, pp.~244--251.
\newblock \href {https://doi.org/10.5555/946250.946946} {\path{doi:10.5555/946250.946946}}.

\bibitem[FHXW22]{Feng2022Learning}
\textsc{Feng X., Huang W., Xu W., Wang H.}:
\newblock Learning-based bending stiffness parameter estimation by a drape tester.
\newblock \emph{ACM TOG 41}, 6 (nov 2022).
\newblock \href {https://doi.org/10.1145/3550454.3555464} {\path{doi:10.1145/3550454.3555464}}.

\bibitem[FSKP23]{Ferguson2023}
\textsc{Ferguson Z., Schneider T., Kaufman D., Panozzo D.}:
\newblock In-timestep remeshing for contacting elastodynamics.
\newblock \emph{ACM TOG 42}, 4 (Jul 2023).
\newblock \href {https://doi.org/10.1145/3592428} {\path{doi:10.1145/3592428}}.

\bibitem[GDP{\etalchar{*}}06]{Grinspun2006SigCourse}
\textsc{Grinspun E., Desbrun M., Polthier K., Schr{\"o}der P., Stern A.}:
\newblock Discrete differential geometry: an applied introduction.
\newblock In \emph{ACM SIGGRAPH 2006 Courses} (2006), SIGGRAPH '06.
\newblock \href {https://doi.org/10.1145/3245634} {\path{doi:10.1145/3245634}}.

\bibitem[GGRZ06]{Grinspun2006ComputingDS}
\textsc{Grinspun E., Gingold Y.~I., Reisman J.~L., Zorin D.}:
\newblock Computing discrete shape operators on general meshes.
\newblock \emph{Computer Graphics Forum 25}, 3 (Sept. 2006), 547--556.
\newblock (Proc. Eurographics'06) \httpsAddr{//dlold.eg.org/handle/10.2312/CGF.v25i3pp547-556}.
\newblock \href {https://doi.org/10.1111/j.1467-8659.2006.00974.x} {\path{doi:10.1111/j.1467-8659.2006.00974.x}}.

\bibitem[GGWZ07]{Garg2007cubicshell}
\textsc{Garg A., Grinspun E., Wardetzky M., Zorin D.}:
\newblock Cubic shells.
\newblock In \emph{Proceedings of the 2007 ACM SIGGRAPH/Eurographics Symposium on Computer Animation} (2007), SCA '07, p.~91–98.
\newblock \href {https://doi.org/10.5555/1272690.1272703} {\path{doi:10.5555/1272690.1272703}}.

\bibitem[GHDS03]{grinspun2003discrete}
\textsc{Grinspun E., Hirani A.~N., Desbrun M., Schr\"{o}der P.}:
\newblock Discrete shells.
\newblock In \emph{Proceedings of the 2003 ACM SIGGRAPH/Eurographics Symposium on Computer Animation} (2003), SCA '03, p.~62–67.
\newblock \href {https://doi.org/10.5555/846276.846284} {\path{doi:10.5555/846276.846284}}.

\bibitem[GKS02]{Grinspun2002CHARMS}
\textsc{Grinspun E., Krysl P., Schr\"{o}der P.}:
\newblock Charms: a simple framework for adaptive simulation.
\newblock \emph{ACM TOG 21}, 3 (Jul 2002), 281–290.
\newblock \href {https://doi.org/10.1145/566654.566578} {\path{doi:10.1145/566654.566578}}.

\bibitem[GSH{\etalchar{*}}04]{gingold2004discrete}
\textsc{Gingold Y., Secord A., Han J.~Y., Grinspun E., Zorin D.}:
\newblock A discrete model for inelastic deformation of thin shells.
\newblock In \emph{Proceedings of the 2004 ACM SIGGRAPH/Eurographics Symposium on Computer Animation} (2004), SCA '04.

\bibitem[GT07]{gardsback2007comparison}
\textsc{G{\"a}rdsback M., Tibert G.}:
\newblock A comparison of rotation-free triangular shell elements for unstructured meshes.
\newblock \emph{Computer Methods in Applied Mechanics and Engineering 196}, 49-52 (Nov 2007), 5001--5015.
\newblock \href {https://doi.org/10.1016/j.cma.2007.06.017} {\path{doi:10.1016/j.cma.2007.06.017}}.

\bibitem[KKB18]{Kugelstadt2018fast}
\textsc{Kugelstadt T., Koschier D., Bender J.}:
\newblock Fast corotated fem using operator splitting.
\newblock In \emph{Proceedings of the 2018 ACM SIGGRAPH/Eurographics Symposium on Computer Animation} (2018), SCA '18.
\newblock \href {https://doi.org/10.1111/cgf.13520} {\path{doi:10.1111/cgf.13520}}.

\bibitem[KMOW00]{kane2000variational}
\textsc{Kane C., Marsden J.~E., Ortiz M., West M.}:
\newblock Variational integrators and the newmark algorithm for conservative and dissipative mechanical systems.
\newblock \emph{International Journal for Numerical Methods in Engineering 49}, 10 (Dec 2000), 1295--1325.
\newblock \href {https://doi.org/10.1002/1097-0207(20001210)49:10<1295::AID-NME993>3.0.CO;2-W} {\path{doi:10.1002/1097-0207(20001210)49:10<1295::AID-NME993>3.0.CO;2-W}}.

\bibitem[LDB{\etalchar{*}}23]{Le2023Second-Order}
\textsc{Le Q., Deng Y., Bu J., Zhu B., Du T.}:
\newblock Second-order finite elements for deformable surfaces.
\newblock In \emph{SIGGRAPH Asia 2023 Conference Papers} (2023).
\newblock \href {https://doi.org/10.1145/3610548.3618186} {\path{doi:10.1145/3610548.3618186}}.

\bibitem[LFFJB24]{loschner2024curved}
\textsc{L{\"o}schner F., Fern{\'a}ndez-Fern{\'a}ndez J.~A., Jeske S.~R., Bender J.}:
\newblock Curved three-director cosserat shells with strong coupling.
\newblock In \emph{Proceedings of the 2024 ACM SIGGRAPH/Eurographics Symposium on Computer Animation} (2024), SCA '24.
\newblock \href {https://doi.org/10.1111/cgf.15183} {\path{doi:10.1111/cgf.15183}}.

\bibitem[LFS{\etalchar{*}}20]{Li2020IPC}
\textsc{Li M., Ferguson Z., Schneider T., Langlois T., Zorin D., Panozzo D., Jiang C., Kaufman D.~M.}:
\newblock Incremental potential contact: Intersection- and inversion-free large deformation dynamics.
\newblock \emph{ACM TOG 39}, 4 (2020).
\newblock \href {https://doi.org/10.1145/3386569.3392425} {\path{doi:10.1145/3386569.3392425}}.

\bibitem[Lia24]{liang2024smoothed}
\textsc{Liang Q.}:
\newblock smoothed hinge model for cloth simulaiton.
\newblock In \emph{Proceedings of the 2024 ACM SIGGRAPH/Eurographics Symposium on Computer Animation} (2024), SCA '24.
\newblock \href {https://doi.org/10.2312/sca.20241161} {\path{doi:10.2312/sca.20241161}}.

\bibitem[LKJ21]{Li2021CIPC}
\textsc{Li M., Kaufman D.~M., Jiang C.}:
\newblock Codimensional incremental potential contact.
\newblock \emph{ACM TOG 40}, 4 (Jul 2021).
\newblock \href {https://doi.org/10.1145/3450626.3459767} {\path{doi:10.1145/3450626.3459767}}.

\bibitem[MDM{\etalchar{*}}02]{Muller2002Stable}
\textsc{M\"{u}ller M., Dorsey J., McMillan L., Jagnow R., Cutler B.}:
\newblock Stable real-time deformations.
\newblock In \emph{Proceedings of the 2002 ACM SIGGRAPH/Eurographics Symposium on Computer Animation} (2002), SCA '02, p.~49–54.
\newblock \href {https://doi.org/10.1145/545261.545269} {\path{doi:10.1145/545261.545269}}.

\bibitem[NSO12]{Narain2012arcsim}
\textsc{Narain R., Samii A., O'Brien J.~F.}:
\newblock Adaptive anisotropic remeshing for cloth simulation.
\newblock \emph{ACM TOG 31}, 6 (nov 2012).
\newblock \href {https://doi.org/10.1145/2366145.2366171} {\path{doi:10.1145/2366145.2366171}}.

\bibitem[OZ00]{onate2000rotation}
\textsc{Onate E., Z{\'a}rate F.}:
\newblock Rotation-free triangular plate and shell elements.
\newblock \emph{International Journal for Numerical Methods in Engineering 47}, 1-3 (Jan 2000), 557--603.
\newblock \href {https://doi.org/10.1002/(SICI)1097-0207(20000110/30)47:1/3<557::AID-NME784>3.0.CO;2-9} {\path{doi:10.1002/(SICI)1097-0207(20000110/30)47:1/3<557::AID-NME784>3.0.CO;2-9}}.

\bibitem[PNdJO14]{Pfaff2014tearing}
\textsc{Pfaff T., Narain R., de~Joya J.~M., O'Brien J.~F.}:
\newblock Adaptive tearing and cracking of thin sheets.
\newblock \emph{ACM TOG 33}, 4 (jul 2014).
\newblock \href {https://doi.org/10.1145/2601097.2601132} {\path{doi:10.1145/2601097.2601132}}.

\bibitem[RGZ07]{reisman2007note}
\textsc{Reisman J., Grinspun E., Zorin D.}:
\newblock \emph{A note on the triangle-centered quadratic interpolation discretization of the shape operator: Technical report}.
\newblock Tech. rep., NYU and Columbia, 2007.

\bibitem[RLR{\etalchar{*}}21]{Romero2021test}
\textsc{Romero V., Ly M., Rasheed A.~H., Charrondi\`{e}re R., Lazarus A., Neukirch S., Bertails-Descoubes F.}:
\newblock Physical validation of simulators in computer graphics: a new framework dedicated to slender elastic structures and frictional contact.
\newblock \emph{ACM TOG 40}, 4 (jul 2021).
\newblock \href {https://doi.org/10.1145/3450626.3459931} {\path{doi:10.1145/3450626.3459931}}.

\bibitem[SLL04]{sze2004popular}
\textsc{Sze K., Liu X., Lo S.}:
\newblock Popular benchmark problems for geometric nonlinear analysis of shells.
\newblock \emph{Finite elements in analysis and design 40}, 11 (July 2004), 1551--1569.
\newblock \href {https://doi.org/10.1016/j.finel.2003.11.001} {\path{doi:10.1016/j.finel.2003.11.001}}.

\bibitem[SZH24]{sauer2024simple}
\textsc{Sauer R.~A., Zou Z., Hughes T.~J.}:
\newblock A simple and efficient hybrid discretization approach to alleviate membrane locking in isogeometric thin shells.
\newblock \emph{Computer Methods in Applied Mechanics and Engineering 424} (May 2024), 116869.
\newblock \href {https://doi.org/10.1016/j.cma.2024.116869} {\path{doi:10.1016/j.cma.2024.116869}}.

\bibitem[Tam13]{Tamstorf2013TecReport}
\textsc{Tamstorf R.}:
\newblock \emph{Derivation of discrete bending forces and their gradients}.
\newblock Tech. rep., Walt Disney Animation Studios, 2013.

\bibitem[TG13]{Tamstorf2013discrete}
\textsc{Tamstorf R., Grinspun E.}:
\newblock Discrete bending forces and their jacobians.
\newblock \emph{Graph. Models 75}, 6 (nov 2013).
\newblock \href {https://doi.org/10.1016/j.gmod.2013.07.001} {\path{doi:10.1016/j.gmod.2013.07.001}}.

\bibitem[TPBF87]{Terzopoulos1987Elastically}
\textsc{Terzopoulos D., Platt J., Barr A., Fleischer K.}:
\newblock Elastically deformable models.
\newblock In \emph{Proc. SIGGRAPH '87} (1987), p.~205–214.
\newblock \href {https://doi.org/10.1145/37401.37427} {\path{doi:10.1145/37401.37427}}.

\bibitem[TWK{\etalchar{*}}59]{timoshenko1959theory}
\textsc{Timoshenko S., Woinowsky-Krieger S., et~al.}:
\newblock \emph{Theory of plates and shells}.
\newblock McGraw-hill New York, 1959.

\bibitem[TWS06]{Thomaszewski2006subdiv}
\textsc{Thomaszewski B., Wacker M., Stra\ss{}er W.}:
\newblock A consistent bending model for cloth simulation with corotational subdivision finite elements.
\newblock In \emph{Proceedings of the 2006 ACM SIGGRAPH/Eurographics Symposium on Computer Animation} (2006), SCA '06, p.~107–116.
\newblock \href {https://doi.org/10.5555/1218064.1218079} {\path{doi:10.5555/1218064.1218079}}.

\bibitem[VCMT95]{Volino1995}
\textsc{Volino P., Courchesne M., Magnenat~Thalmann N.}:
\newblock Versatile and efficient techniques for simulating cloth and other deformable objects.
\newblock In \emph{Proc. SIGGRAPH '95} (1995), p.~137–144.
\newblock \href {https://doi.org/10.1145/218380.218432} {\path{doi:10.1145/218380.218432}}.

\bibitem[WB23]{Wen2023KLShell}
\textsc{Wen J., Barbi\v{c} J.}:
\newblock Kirchhoff-love shells with arbitrary hyperelastic materials.
\newblock \emph{ACM TOG 42}, 6 (Dec 2023).
\newblock \href {https://doi.org/10.1145/3618405} {\path{doi:10.1145/3618405}}.

\bibitem[WBH{\etalchar{*}}07]{WARDETZKY2007}
\textsc{Wardetzky M., Bergou M., Harmon D., Zorin D., Grinspun E.}:
\newblock Discrete quadratic curvature energies.
\newblock \emph{Computer Aided Geometric Design 24}, 8 (2007), 499--518.
\newblock \href {https://doi.org/https://doi.org/10.1016/j.cagd.2007.07.006} {\path{doi:https://doi.org/10.1016/j.cagd.2007.07.006}}.

\bibitem[Wei12]{weischedel2012discrete}
\textsc{Weischedel C.}:
\newblock \emph{A discrete geometric view on shear-deformable shell models}.
\newblock Georg-August-Universität Göttingen., 2012.

\bibitem[Zor05]{Zorin2005Curvature}
\textsc{Zorin D.}:
\newblock Curvature-based energy for simulation and variational modeling.
\newblock In \emph{Proceedings of the International Conference on Shape Modeling and Applications 2005} (2005), SMI '05, p.~198–206.
\newblock \href {https://doi.org/10.1109/SMI.2005.14} {\path{doi:10.1109/SMI.2005.14}}.

\bibitem[ZS12]{zhou2012geometric}
\textsc{Zhou Y., Sze K.~Y.}:
\newblock A geometric nonlinear rotation-free triangle and its application to drape simulation.
\newblock \emph{International journal for numerical methods in engineering 89}, 4 (Jan 2012), 509--536.
\newblock \href {https://doi.org/10.1002/nme.3250} {\path{doi:10.1002/nme.3250}}.

\bibitem[ZSTB10]{Zhu2010multigrid}
\textsc{Zhu Y., Sifakis E., Teran J., Brandt A.}:
\newblock An efficient multigrid method for the simulation of high-resolution elastic solids.
\newblock \emph{ACM TOG 29}, 2 (Apr 2010).
\newblock \href {https://doi.org/10.1145/1731047.1731054} {\path{doi:10.1145/1731047.1731054}}.

\end{thebibliography}
\appendix
\section{Corotational Transformation} \label{appendixA} 
Consider a point $\tilde{\mathbf{x}}$ on the shell stencil in the current corotational frame with origin at at $\mathbf{x}_{{o}}$. The point $\tilde{\mathbf{x}}$ is related to its counterpart $\mathbf{x}$ in the world frame by the transformation
\begin{equation} \label{eq65}
    \tilde{\mathbf{x}} = \begin{bmatrix}\mathbf{n}_{\tilde{x}} & \mathbf{n}_{\tilde{y}} & \mathbf{n}_{\tilde{z}}\end{bmatrix}^{{T}}(\mathbf{x} - \mathbf{x}_{{o}}),
\end{equation}
where $\mathbf{n}_{\tilde{x}}$, $\mathbf{n}_{\tilde{y}}$, and $\mathbf{n}_{\tilde{z}}$ define the directions of the local coordinate axes.
The triangle-centered and edge-based stencils respectively take $\mathbf{x}_1$ and $\mathbf{x}_2$ as their origins.
For the triangle-centered stencil, the $\tilde{x}$-axis direction is defined as $\mathbf{n}_{\tilde{x}} = \mathbf{x}_{12} / \lVert \mathbf{x}_{12} \rVert$ (see Figures~\ref{fig:corotational_FVM_hinge} and~\ref{fig:corotational_smoothed_hinge}), while for the edge-based stencil, it is defined as $\mathbf{n}_{\tilde{x}} = \mathbf{x}_{23} / \lVert \mathbf{x}_{23} \rVert$ (see Figure~\ref{fig:corotational_edge_based_hinge}).
The $\tilde{z}$-axis direction $\mathbf{n}_{\tilde{z}}$ is computed using Eq. (\ref{eq8}) for the edge stencil and Eq. (\ref{eq15}) for the triangle-centered stencil. 
The $\tilde{y}$-axis direction, $\mathbf{n}_{\tilde{y}} = \mathbf{n}_{\tilde{z}} \times \mathbf{n}_{\tilde{x}}$, is determined to according to the right-hand rule.
Similarly, points in the initial corotational frame can be transformed using the same approach.

\section{Gradient of Normal Direction of Corotational Shell Stencil} \label{appendixB} 
The direction of $\tilde{z}$-axis, which corresponds to the normal direction of the corotational shell stencil, is given by
\begin{equation} \label{eq66}
    \mathbf{n}_{\tilde{z}} = \frac{\mathbf{n}_{c}}{\lVert \mathbf{n}_{c} \rVert},
\end{equation}
where $\mathbf{n}_{c}$ is the unnormalized normal vector. In this appendix, the gradient operator is denoted by $\nabla = {\partial}/{\partial (\mathbf{x}^{e})^{T}}$. The gradient of the $\tilde{z}$-axis direction is then expressed as
\begin{equation} \label{eq67}
    \nabla \mathbf{n}_{\tilde{z}}=({\mathbf{I}-\mathbf{n}}_{\tilde{z}} \mathbf{n}_{\tilde{z}}^{T})\frac{\nabla \mathbf{n}_{c}}{\lVert \mathbf{n}_{c} \rVert}.
\end{equation}

For the corotational edge-based hinge model, let $\mathbf{p} = \mathbf{x}_{P1}$, $\mathbf{q} = \mathbf{x}_{Q4}$ as the triangle height vectors, and $\mathbf{e} = \mathbf{x}_{23}$ as the hinge edge vector. The gradient of $\mathbf{n}_{c}$ is given by
\begin{equation} \label{eq68}
    \nabla\mathbf{n}_{c}^{}=\nabla (\frac{\mathbf{p}}{\lVert\mathbf{p}\rVert})+\nabla (\frac{\mathbf{q}}{\lVert\mathbf{q}\rVert}),
\end{equation}
where the gradient of the triangle height directions are
\begin{equation} \label{eq69}
    \nabla (\frac{\mathbf{p}}{\lVert\mathbf{p}\rVert})=(\frac{\lVert\mathbf{p}\rVert \mathbf{I}-\mathbf{p}\mathbf{p}^{T}}{\lVert\mathbf{p}\rVert^{3}})\nabla\mathbf{p},
    \text{and} \ \nabla (\frac{\mathbf{q}}{\lVert\mathbf{q}\rVert})=(\frac{\lVert\mathbf{q}\rVert\mathbf{I}-\mathbf{q}\mathbf{q}^{T}}{\lVert\mathbf{q}\rVert^{3}})\nabla\mathbf{q}.
\end{equation}
The components of $\nabla \mathbf{p}$ are
\begin{equation} \label{eq70}
    \begin{gathered}
    \frac{\partial \mathbf{p}}{\partial (\mathbf{x}_1)^{T}}=\mathbf{I}-\frac{\mathbf{e} \mathbf{e}^T}{\mathbf{e}^{2}}, 
    \frac{\partial \mathbf{p}}{\partial (\mathbf{x}_2)^{T}}=-\left(1-s_p\right)\left(\mathbf{I}-\frac{\mathbf{e} \mathbf{e}^T}{\mathbf{e}^{2}}\right)+\frac{\mathbf{e} \mathbf{p}^T}{\mathbf{e}^{2}}, \\
    \frac{\partial \mathbf{p}}{\partial (\mathbf{x}_3)^{T}}=-s_p\left(\mathbf{I}-\frac{\mathbf{e} \mathbf{e}^T}{\mathbf{e}^{2}}\right)-\frac{\mathbf{e} \mathbf{p}^T}{\mathbf{e}^{2}},
    \frac{\partial \mathbf{p}}{\partial (\mathbf{x}_4)^{T}}=\mathbf{0},
    \end{gathered}
\end{equation}
and the components of $\nabla \mathbf{q}$ are
\begin{equation} \label{eq71}
    \begin{gathered}
    \frac{\partial \mathbf{q}}{\partial (\mathbf{x}_1)^{T}}=\mathbf{0},
    \frac{\partial \mathbf{q}}{\partial (\mathbf{x}_2)^{T}}=-\left(1-s_q\right)\left(\mathbf{I}-\frac{\mathbf{e} \mathbf{e}^T}{\mathbf{e}^{2}}\right)+\frac{\mathbf{e} \mathbf{q}^T}{\mathbf{e}^{2}}, \\
    \frac{\partial \mathbf{q}}{\partial (\mathbf{x}_3)^{T}}=-s_q\left(\mathbf{I}-\frac{\mathbf{e} \mathbf{e}^T}{\mathbf{e}^{2}}\right)-\frac{\mathbf{e} \mathbf{q}^T}{\mathbf{e}^{2}},
    \frac{\partial \mathbf{q}}{\partial (\mathbf{x}_4)^{T}}=\mathbf{I}-\frac{\mathbf{e} \mathbf{e}^T}{\mathbf{e}^{2}}.
    \end{gathered}
\end{equation}
Here, the coefficients $s_p= \mathbf{x}_{21} \cdot \mathbf{x}_{23}/{\mathbf{x}_{23}^2}$ and $s_q= \mathbf{x}_{24} \cdot \mathbf{x}_{23}/{\mathbf{x}_{23}^2}$ are defined accordingly.

For the corotational FVM hinge and smoothed hinge bending model, the normal $\mathbf{n}_c$ is given by
\begin{equation} \label{eq72}
    \mathbf{n}_c = \mathbf{x}_{12} \times \mathbf{x}_{13},
\end{equation}
so the components of $\nabla \mathbf{n}_c$ are
\begin{equation} \label{eq73}
    \begin{gathered}
    \frac{\partial \mathbf{n}_c}{\partial (\mathbf{x}_1)^{T}}=(\mathbf{x}_3 - \mathbf{x}_2)^{\times},
    \frac{\partial \mathbf{n}_c}{\partial (\mathbf{x}_3)^{T}}=(\mathbf{x}_2 - \mathbf{x}_1)^{\times},
    \frac{\partial \mathbf{n}_c}{\partial (\mathbf{x}_5)^{T}}=\mathbf{0},\\
    \frac{\partial \mathbf{n}_c}{\partial (\mathbf{x}_2)^{T}}=(\mathbf{x}_1 - \mathbf{x}_3)^{\times},
    \frac{\partial \mathbf{n}_c}{\partial (\mathbf{x}_4)^{T}}=\mathbf{0},
    \frac{\partial \mathbf{n}_c}{\partial (\mathbf{x}_6)^{T}}=\mathbf{0},
    \end{gathered}
\end{equation}
where the operator $\times$ applied to a vector $\mathbf{v} = \begin{bmatrix} x&y&z\end{bmatrix}^{T}$ yields the skew-symmetric matrix
\begin{equation} \label{eq74}
    \mathbf{v}^{\times} = 
    \begin{bmatrix}
          0 & -z & y \\
          z & 0 & -x \\
          -y & x & 0
    \end{bmatrix},
\end{equation}
which corresponds to the cross product $\mathbf{v}\times(*)$.

Finally, by vectorizing the $(\nabla \mathbf{n}_c)^{T}$, we obtain
\begin{equation} \label{eq75}
    \frac{\partial \mathbf{n}_{\tilde{z}}}{\partial \mathbf{x}^{e}} = vec\left((\nabla \mathbf{n}_c\right)^{T}).
\end{equation}

\begin{figure}
  \centering
  \includegraphics[width=.5\linewidth]{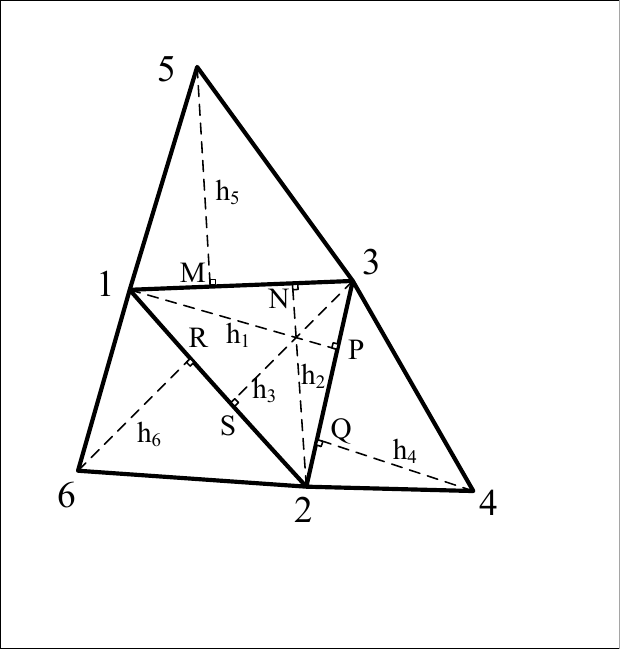}
  \caption{\label{fig:geo_quan}
     Definition of geometric quantities. The heights of the triangles are denoted as $h_i$ where $i$ ranges from 1 to 6, and the perpendicular feet are represented by $M, N, P, Q, R, and S$.
   }
\end{figure}

\section{The Directional Curvature Operator} \label{appendixC} 
The directional curvature operator can be quantified using the geometry quantities in Figure~\ref{fig:geo_quan}, as follow
\begin{equation} \label{eq76}
    \mathbf{L}_{p} = -\mathbf{H} \boldsymbol{\theta},
\end{equation}
where 
\begin{equation} \label{eq77}
    \mathbf{H}=\begin{bmatrix}\frac{2}{h_{1}+h_{4}}&0&0\\ 0&\frac{2}{h_{2}+h_{5}}&0\\   0&0&\frac{2}{h_{3}+h_{6}}\end{bmatrix},
\end{equation}
and $\boldsymbol{\theta}$ collects the linearized bend angles as follow
\begin{equation} \label{eq78}
    \begin{bmatrix}-\frac{1}{h_{1}} & \left(\frac{\lVert \tilde{\mathbf{X}}_{M3} \rVert}{ah_{5}} + \frac{\lVert \tilde{\mathbf{X}}_{N3}\rVert}{ah_{2}}\right) & \left(\frac{\lVert \tilde{\mathbf{X}}_{R2}\rVert}{bh_{6}} + \frac{\lVert \tilde{\mathbf{X}}_{S2}\rVert}{bh_{3}}\right) \\
    \left(\frac{\lVert \tilde{\mathbf{X}}_{Q3}\rVert}{c h_{4}} + \frac{\lVert \tilde{\mathbf{X}}_{P3}\rVert}{c h_{1}}\right) & -\frac{1}{h_{2}} & \left(\frac{\lVert \tilde{\mathbf{X}}_{R1}\rVert}{bh_{6}} + \frac{\lVert \tilde{\mathbf{X}}_{S1}\rVert}{b h_{3}}\right) \\
    \left(\frac{\lVert \tilde{\mathbf{X}}_{Q2}\rVert}{ch_{4}} + \frac{\lVert \tilde{\mathbf{X}}_{P2} \rVert}{c h_{1}}\right) & \left(\frac{\lVert \tilde{\mathbf{X}}_{M1}\rVert}{ah_{5}} + \frac{\lVert \tilde{\mathbf{X}}_{N1} \rVert}{a h_{2}}\right) & -\frac{1}{h_{3}} \\-\frac{1}{h_{4}} & 0 & 0 \\0 & -\frac{1}{h_{5}} & 0 \\0 & 0 & -\frac{1}{h_{6}}\end{bmatrix}^{T},      
\end{equation}
with $a=\lVert \tilde{\mathbf{X}}_{31}\rVert$, $b=\lVert \tilde{\mathbf{X}}_{12}\rVert$ and $c=\lVert \tilde{\mathbf{X}}_{23}\rVert$.

\section{Quadratic Thin Plate/Shell and Accuracy Discrepancy} \label{appendixD} 
In this Appendix, the terminologies come from the Section~\ref{Corotational_edge-based_hinge_bending_model}.

\noindent\textit{Quadratic Thin Plate (QTP).} When the directional curvature operator is quantified in the world frame, the bending energy of our EP model in Eq. (\ref{eq20}) can be rewritten as  
\begin{equation} \label{eq79}
    \Psi_b^{QTP}=\frac{A_{\mathcal{E}}}{2} k_b (\mathbf{x}^{e})^{T}(\mathbf{L}_{n}^{T}\mathbf{L}_{n}\otimes\mathbf{I}) \mathbf{x}^{e},
\end{equation}
where the curvature operator $\mathbf{L}_{n}$ is
\begin{equation} 
    c_h\begin{bmatrix}
    \frac{1}{h_{1}}  & -\left( \frac{\lVert {\mathbf{X}}_{P3}\rVert}{\lVert {\mathbf{X}}_{23}\rVert h_{1}} + \frac{\lVert {\mathbf{X}}_{Q3}\rVert}{\lVert {\mathbf{X}}_{23}\rVert h_{4}} \right) &  -\left( \frac{\lVert {\mathbf{X}}_{P2}\rVert}{\lVert {\mathbf{X}}_{23}\rVert h_{1}} + \frac{\lVert {\mathbf{X}}_{Q2}\rVert}{\lVert {\mathbf{X}}_{23}\rVert h_{4}} \right) &  \frac{1}{h_{4}}  
    \end{bmatrix}
\end{equation}
with $c_h = 2 / (h_1 + h_4)$.
The Hessian of the QTP's bending energy is
\begin{equation} 
    \frac{\partial^{2} \Psi_b^{QTP}}{\partial \mathbf{x}^{e}{\partial (\mathbf{x}^{e})^{T}}} = k_b A_{\mathcal{E}}  \mathbf{L}_{n}^{T}\mathbf{L}_{n}\otimes\mathbf{I},
\end{equation}
and the gradient of the QTP's bending energy is
\begin{equation} 
    \frac{\partial  \Psi_b^{QTP}}{\partial \mathbf{x}^{e}}= (\frac{\partial^{2} \Psi_b^{QTP}}{\partial \mathbf{x}^{e}{\partial (\mathbf{x}^{e})^{T}}}) \mathbf{x}^{e}.
\end{equation} 

\noindent\textit{Quadratic Thin Shell (QTS).} The curvature operator in Eq. (\ref{eq79}) can also be extended to the bending energy of our ES model. The bending energy of QTS is
\begin{equation} 
    \Psi_b^{QTS} = \frac{1}{2} A_{\mathcal{E}} k_b \epsilon_b^{2},
\end{equation}
where the curvature change $\epsilon_b$ is
\begin{equation} 
    \epsilon_b = {\kappa}_m-({\kappa}_{m})_{0}=\mathbf{L}_{n}\mathbf{N}^{T} {\mathbf{x}}^{e}-\mathbf{L}_{n}\mathbf{N}_{0}^{T} {\mathbf{X}}^{e}.
\end{equation}
The gradient of the QTS's bending energy is 
\begin{equation} 
    \frac{\partial \Psi_b^{ES}}{\partial \mathbf{x}^{e}}={A_{\mathcal{E}}} k_b  \frac{\partial \epsilon_b}{\partial \mathbf{x}^{e}}\epsilon_b,
\end{equation}
where the gradient of the curvature change can be generalized from Eq. (\ref{eq26}), i.e.,
\begin{equation} 
    \frac{\partial \epsilon_b}{\partial \mathbf{x}^{e}} = ({\mathbf{x}}^{e})^{T}\frac{\partial \mathbf{N}}{\partial \mathbf{x}^{e}} \mathbf{L}_{n}^{T}+\mathbf{N}\mathbf{L}_{n}^{T}.
\end{equation}
The Hessian of QTS' bending energy is the same as the Hessian of QTP'.
QTP/QTS can be seen as a variation of our EP/ES that quantifies the curvature operator in the world frame.

\noindent\textit{Accuracy Discrepancy.} In comparison, the Quadratic Shell model~\cite{Bergou2006Aquadratic} is the linearized version of the Discrete Shell model~\cite{grinspun2003discrete} of which detailed numerical formulations and isotropic bending rigidity can be found in~\cite{Tamstorf2013discrete, Tamstorf2013TecReport}. By expressing the discretized bending energy in Eq. (\ref{eq79}) using the terminologies from the Quadratic Shell paper~\cite{Bergou2006Aquadratic}, we can conclude that the bending energy of our EP/QTP model is three times less than the bending energy of Quadratic Shell. 
The Cubic Shell model~\cite{Garg2007cubicshell} builds on a foundation laid out from Quadratic Shell~\cite{Bergou2006Aquadratic}, so the accuracy discrepancy holds. Also, the Quadratic Shell can be seen as a rest-flat version of the Cubic Shell.

To account for the accuracy discrepancy of our edge-stenciled models (EP/QTP/ES/QTS) with Quadratic/Cubic Shells~\cite{ Bergou2006Aquadratic, Garg2007cubicshell}, we example the bending energy of thin plate model we computed from Eq. (\ref{eq16}), i.e.,
\begin{equation} \label{eq87}
    \Psi_b = \frac{1}{2} A_{\mathcal{E}} k_b \kappa_{{m}}^{2}.
\end{equation}
Under the small strain/curvature assumption, the directional curvature $\kappa_{{m}}$ in Eq. (\ref{eq4}) is
\begin{equation} 
    \kappa_{{m}} = \frac{2\theta}{h_1 + h_4} = \frac{\theta l}{A_1 + A_4}, 
\end{equation}
and $A_{\mathcal{E}}=A_1 + A_4$ is consistent with the area on which the curvature $\kappa_{{m}}$ is defined. $A_1$ and $A_4$ are the triangle areas of one edge stencil in the initial configuration. $l$ is the length of the hinge edge. By using our terminologies, the underlined bending energy of Quadratic/Cubic Shells from~\cite{grinspun2003discrete, Tamstorf2013discrete, Tamstorf2013TecReport} is
\begin{equation} \label{eq89}
    \Psi_b^{'} = \frac{1}{2} A_{\mathcal{E}}^{'} k_b (\kappa_{{m}}^{'})^{2}.
\end{equation}
Here, 
\begin{equation} 
    \kappa_{{m}}^{'} = \frac{\theta l}{\frac{1}{3}(A_1 + A_4)}=3\frac{2\theta}{h_1 + h_4}=3\kappa_{{m}}, 
\end{equation}
and the $A_{\mathcal{E}}^{'}={(A_1 + A_4)}/{3}=A_{\mathcal{E}}/3$ is consistent with the area on which the curvature $\kappa_{{m}}^{'}$ is defined. By comparing Eq. (\ref{eq87}) and Eq. (\ref{eq89}), we can obtain 
\begin{equation} 
    \Psi_b^{'} = 3\Psi_b.
\end{equation}
Thus, scaling down the Quadratic/Cubic Shell’s bending energy allows the Quadratic/Cubic Shell to yield accurate predictions in the linear (small deflection) plate bending benchmark (Section~\ref{Linear_plate_bending_benchmark}).

\end{document}